\documentclass[%
 reprint,
superscriptaddress,
groupedaddress,
 amsmath,amssymb,
 aps,
]{revtex4-2}

\usepackage{graphicx}
\usepackage{dcolumn}
\usepackage{bm}
\usepackage{epstopdf}
\usepackage{color}
\usepackage{amsmath}
\usepackage{braket}
 \usepackage{amssymb}
 \usepackage{amsfonts}
\usepackage{changes}
\usepackage[colorlinks=true,linkcolor=blue,urlcolor=blue,citecolor=blue]{hyperref}

\begin{document}

\preprint{APS/123-QED}

\title{Intrinsic chiral field as vector potential of the magnetic current in the zig-zag lattice of magnetic dipoles.}%

\author{Paula Mellado}
\affiliation{School of Engineering and Sciences, 
	Universidad Adolfo Ib{\'a}{\~n}ez,
	Santiago, Chile}
	
\author{Kevin Hofhuis}
 \affiliation{Laboratory for Mesoscopic Systems, Department of Materials, ETH Zurich, Switzerland}
 \affiliation{
 Laboratory for Multiscale Materials Experiments, Paul Scherrer Institute, Switzerland
}
 \affiliation{Department of Applied Physics, Yale University, New Haven, USA}
	
\author{Ignacio Tapia}
\affiliation{Faculty of Sciences,   
	Universidad de Chile,
	Santiago, Chile}
	
\author{Andres Concha}
\affiliation{School of Engineering and Sciences, 
	Universidad Adolfo Ib{\'a}{\~n}ez,
	Santiago, Chile}

\date{\today}

\begin{abstract}
Chiral magnetic insulators manifest novel phases of matter where the sense of rotation of the magnetization is associated with exotic transport phenomena. Effective control of such phases and their dynamical evolution points to the search and study of chiral fields like the Dzyaloshinskii-Moriya interaction. 
Here we combine experiments, numerics, and theory to study a zig-zag dipolar lattice as a model of an interface between magnetic in-plane layers with a perpendicular magnetization. The zig-zag lattice comprises two parallel sublattices of dipoles with perpendicular easy plane of rotation. The dipolar energy of the system is exactly separable into a sum of symmetric and antisymmetric long-range exchange interactions between dipoles, where the antisymmetric coupling generates a nonlocal Dzyaloshinskii-Moriya field which stabilizes winding textures with the form of chiral solitons. The Dzyaloshinskii-Moriya interaction acts as a vector potential or gauge field of the magnetic current and gives rise to emergent magnetic and electric fields that allow the manifestation of the magnetoelectric effect in the system. 
\end{abstract}

\maketitle

\begin{figure*}
\includegraphics[width=\textwidth]{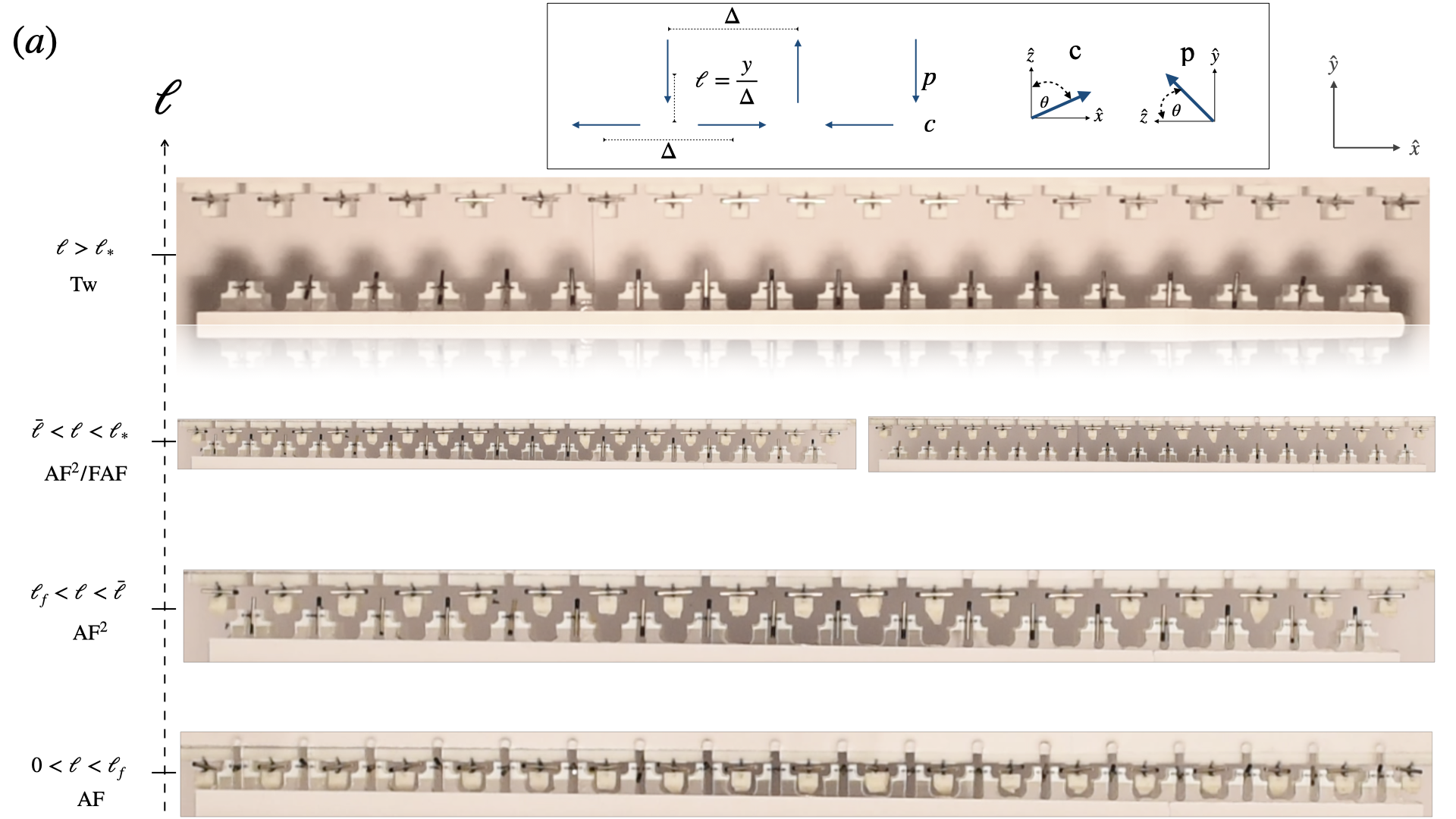}
\includegraphics[width=\textwidth]{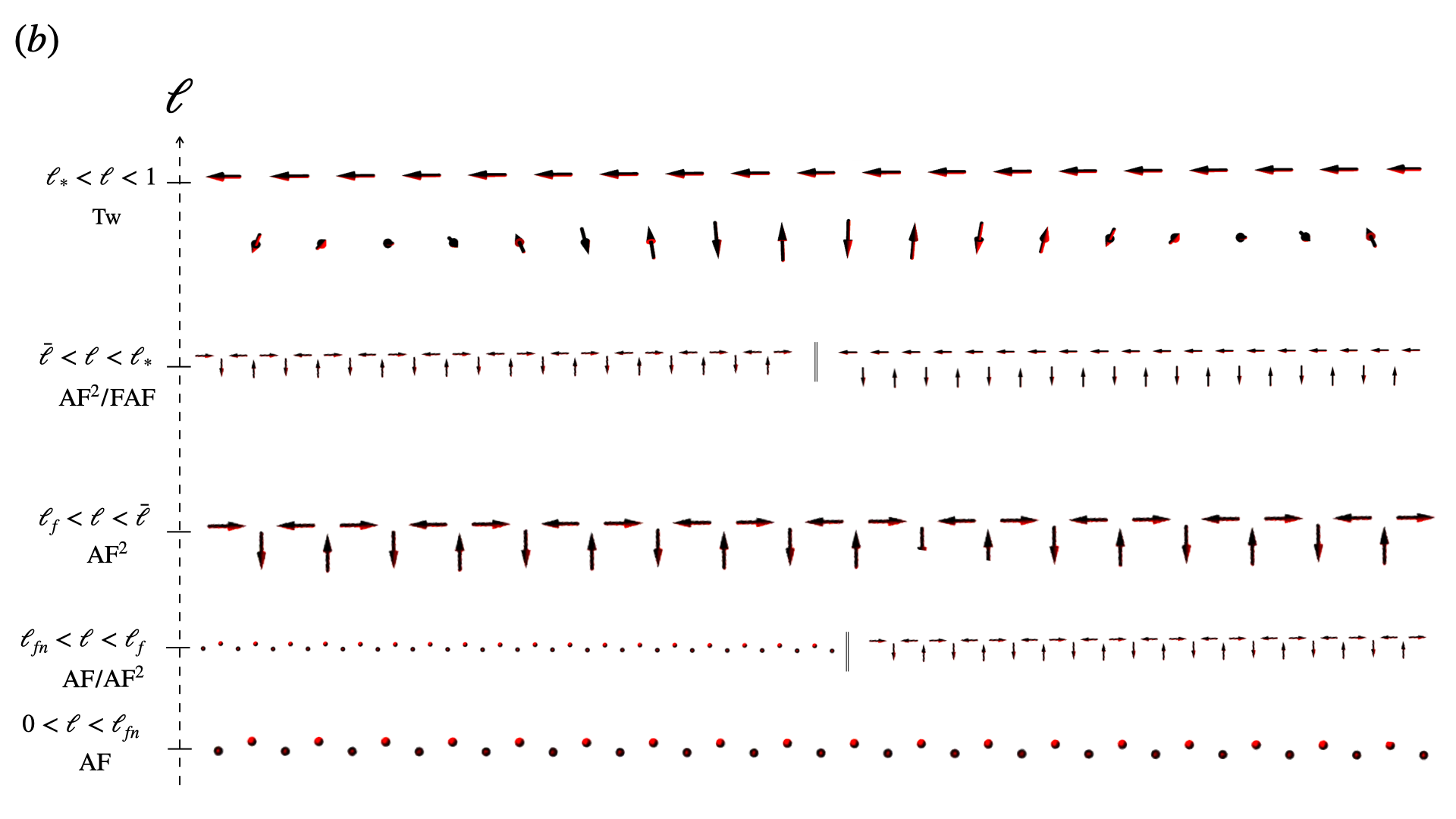}
\caption{(a) Magnetic phases of the experimental zig-zag lattice in terms of $\ell$. From top to bottom as $\ell$ decreases: at large $\ell>\ell_*$, the system settles in the Tw phase. In the range $(\bar{\ell},\ell_*)$ the lattice enters the metastable regime in the $x-y$ plane $\rm AF^2/FAF$.  As  $\ell$ decreases further,  in the range  $\ell\in (\ell_f,\bar{\ell})$, the $\rm AF^2$ phase is selected. At very small gaps $\ell<\ell_f$ dipoles of both sublattices configure an antiferromagnetic state along the  $\hat{z}$ axis. Inset:  The zig-zag lattice's geometry consists of n=37 Neodymium magnets hinged on top of a PTFE plate. All have length $a=12,7\times 10^{-3}$ [m], radius 
$r = 0.79\times 10^{-3}$ [m], mass $0.189\times 10^{-3}$  [kg], and saturation magnetization $M_{s} =1.05\times 10^{6}$ [A/m]. The distance between two consecutive rods in the same chain is $\Delta=22\times 10^{-3}$ m fixed,  and the tunable vertical interchain gap is $\ell=y/\Delta$ where y is the vertical distance measured from chain \emph{c}. Dipoles at chain \emph{c} and \emph{p} rotate in the planes $x-z$ and $y-z$ respectively. (b) Screenshots of the lattice from Molecular dynamics simulations.} 
\label{f1}
\end{figure*}
\begin{figure}
\includegraphics[width=\columnwidth]{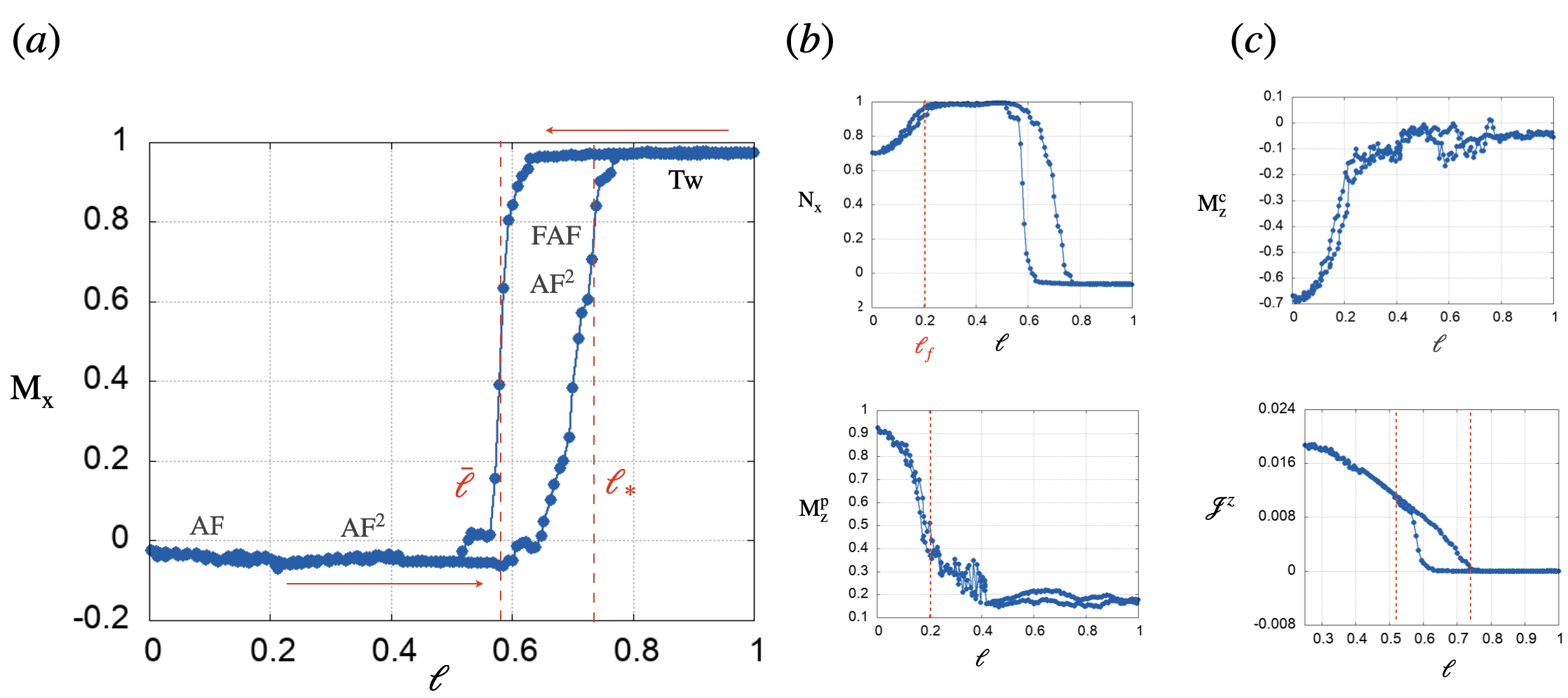}
\caption{Magnetization dynamics in experiments. (a) evolution of $\rm M_x$ and (b) (top panel ) the staggered magnetization along $\hat{x}$, $\rm N_{x}$ in terms of $\ell$. Red arrows pointing to the right and left denote chain \emph{p}  moving apart from and approaching chain \emph{c} respectively. The bottom panel of (b) and the upper panel of (c) show the $\hat{z}$ component of the magnetization of the \emph{p} and \emph{c} sublattices, respectively. The magnetic current along $\hat{z}$ in the metastable regime at intermediate $\ell$ is shown in the bottom panel of figure (c).}
\label{f2}
\end{figure}

\section{\label{sec:intro}Introduction}
Chiral symmetry refers to symmetry under mirror reflection: a subject is said to be chiral when it lacks such symmetry. Chiral asymmetry is rather common in nature at several scales: at the microscale, it is well-known that elementary particles and organic molecules have a preferential chiral characterization, while very large macroscopic systems like spiral galaxies are chiral too \cite{jiang2017fluorescent}. In condensed matter physics, a theory, symmetry, or field is chiral if it is not invariant under the inversion of one spatial dimension. Chiral condensed matter systems realize exotic electronic \cite{takane2019observation}, topological \cite{hasan2021weyl} and magnetic phases \cite{shekhar2018chirality,monteiro2015magnetotransport,li2016chiral,cortijo2016strain,kharzeev2021chiral,xia1997noncollinear,meckler2009real,von2014interface} and structures \cite{komineas2015skyrmion,yang2021chiral}. The transport properties of chiral matter can undergo quantum anomalies associated with chiral symmetry as in the topological Dirac/Weyl semi-metals \cite{liang2018experimental}, and chiral interactions and fields play roles as remarkable as the manifestation of natural optical activity in materials that lack mirror symmetry \cite{claborn2008optical}. 
The antisymmetric Dzyaloshinskii-Moriya interaction \cite{moriya1960new}, (DMI), is a chiral coupling \cite{ryu2013chiral} able to trigger magnetic torques in magnetic systems. Such torques can stabilize localized and spatially modulated structures with a fixed sense of rotation \cite{dzyaloshinskii1964theory} and influence the transport in such systems; in consequence, they are key to the development of spin-based memory, logic, and signal transmission devices \cite{yang2021chiral}. DMI is generally described by a vector product formed by the localized magnetic moments of two magnetic ions ${\rm U}_{\rm DM}=\bm{D}_{ij}\cdot(\bm{m}_i\times\bm{m}_j)$, where $\bm{D}_{ij}$ is called Dzyaloshinskii-Moriya (DM) vector. Dzyaloshinskii \cite{dzyaloshinsky1958thermodynamic} introduced ${\rm U}_{\rm DM}$ based on phenomenological considerations to explain the observation of weak ferromagnetism in some antiferromagnets. Later, Moriya \cite{moriya1960anisotropic} demonstrated that in low symmetry magnetic crystals, the spin-orbit interaction can lead to DMI by taking into account the spin-orbit coupling (SOC) in the theory of superexchange interaction proposed by Anderson \cite{anderson1950antiferromagnetism}. In this theory $\bm{D}_{ij}$ is proportional to the spin-orbit interaction and depends on the symmetry of crystals. Afterward, it was shown that this mechanism is relevant only when the local symmetry is sufficiently low and that a weak ferromagnetic moment emerges from the superexchange coupling only when more than a single bond is considered \cite{shekhtman1992moriya}. 
Chiral magnetic couplings have been induced in centrosymmetric crystals by symmetry breaking due to electric currents \cite{freimuth2020dynamical}, applied magnetic and electric fields and by strain \cite{deger2020strain}. For bulk magnetic materials, such coupling is generally weak; however, in small artificial structures such as ferromagnetic films, multilayers, nanowires, and nanodots, this is not the case \cite{chen2021observation,fernandez2019symmetry,crepieux1998dzyaloshinsky,heide2008dzyaloshinskii,torrejon2014interface,belmeguenai2015interfacial,gross2016direct,turov1996symmetry}. In two-dimensional films, the interfacial DMI defines a rotational sense for the magnetization which can be used to create chiral magnetic structures like spin-spirals, domain walls, and skyrmions \cite{drchal1996interlayer, alamdar2021domain}. Recently, a strong interlayer Dzyaloshinskii-Moriya interaction has been demonstrated in FM/Pt/FM trilayers with orthogonal magnetization. In this system, the DMI causes a chiral interlayer coupling \cite{bogdanov2001chiral,yang2015anatomy,belabbes2016hund} that favors one-handed orthogonal magnetic configurations of Co and TbFe, as revealed through the Hall effect and magnetoresistance measurements \cite{avci2021chiral}. At the nanoscale, the intralayer DMI has been used to engineer strong, localized intrinsic chiral torques that trigger the spontaneous motion of domain walls or bias the speed of current-driven domain walls in the magnetic race- tracks \cite{liu2021engineering,PhysRevLett.57.2442,lau2016spin,avci2019interface,avci2019interface}. In thin-film metallic systems, spin-orbit coupling arises from a proximal heavy metal, \cite{manchon2009theory} where the metallic layer typically provides the spin-orbit coupling to induce the DMI. 
However, recent experiments in the rare earth garnets \cite{caretta2020interfacial} suggest that a proximate high-SOC layer is not required, and the magnetic ion in the magnetic film itself provides the critical SOC responsible for DMI, irrespective of the SOC of the interfacing material. Indeed, in perpendicularly-magnetized iron garnets, rare-earth orbital magnetism has given rise to an intrinsic spin-orbit coupling generating interfacial DMI at mirror symmetry-breaking interfaces \cite{brock2021dynamic}. Moreover, recent findings showing that the rare-earth ion substitution and strain engineering can significantly alter the DMI \cite{caretta2020interfacial,ding2019interfacial},  remain to be understood. 
\section{\label{sec:sum}Summary of results}
Aimed to identify new sources of chiral fields and stable chiral structures, we study  the magnetization dynamics of macroscopic zig-zag lattices of dipoles with experiments and theory. Magnetic dipoles in different sublattices have perpendicular easy planes of rotation, which, combined with dipolar interactions, allows the exact mapping of the magnetic energy into four energetic contributions, which include symmetric and antisymmetric or chiral long-range interactions between the dipoles.  Dynamics in the system is induced by tuning the chiral torques through the variation of the gap $\ell$, which is the distance between the two sublattices along axis $\hat{y}$ (Fig.\ref{f1}). As $\ell$ is varied, the system transits between four magnetic phases through a rich dynamical process that features hysteresis and stabilizes chiral magnetic textures. The explicit formula for symmetric and chiral couplings reveals the underlying mechanism through which the DM vector acts as a vector potential of an out-of-plane spin current and prompts the onset of magnetic and electric fields. The emergence of a U(1) gauge theory in this system exposes an electric polarization \cite{PhysRevLett.57.2442,lau2016spin,avci2021chiral}.

The paper is organized as follows. In section \ref{sec:model} we present the model system and show the separation of the total energy of it into symmetric and antisymmetric contributions. Section \ref{sec:phases} shows the magnetic phases realized by the lattice as $\ell$ is tuned. In section \ref{sec:couplings} we discuss the contributions of the effective symmetric and antisymmetric couplings to the energetics of the system, and in section \ref{sec:current} we define the magnetic current and torque in terms of the chiral field that arises product of the dipolar interactions.   Section \ref{sec:planar} is devoted to studying the magnetic current, the associated potential vector, and the emergent fields that arise in the planar phase. Section \ref{sec:solo} focuses on the magnetic phase realized at large $\ell$ and the onset of magnetic solitons. Concluding remarks are presented in section \ref{sec:conclusion}.
\section{\label{sec:model}Model}
The magnetic dipolar energy for the system of $n$ dipoles (in units of Joule) in the zig-zag lattice reads 
$
U=\frac{g}{2} \sum_{i\neq k=1}^n \frac{\hat {\bm  m}_i \cdot\hat {\bm m}_k - 
3 (\hat {\bm m}_i \cdot \hat {\bm{e}}_{ik} )(\hat {\bm m}_k\cdot \hat {\bm{e}}_{ik} )}{|{\bm r}_i -{\bm r}_k |^3},
$
where $\hat {\bm{e}}_{ik}= ({\bm r}_i -{\bm r}_k ) /|{\bm r}_i -{\bm r}_k |$, and $g =\frac{\mu_0 m_0^2}{4\pi\Delta^3}$  sets the energy scale. It contains the physical parameters involved in the energy, such as $\Delta$, the lattice constant, $\mu_0$, the magnetic permeability, and $m_0$ (in units $\rm[m^2 A]$), the intensity of the magnetic moments with saturation magnetization $\rm M_s$. Hereafter we normalize all distances by $\Delta$. The magnetic moments are normalized by  $m_0$ and dipoles belonging to sublattice $\alpha:(\emph{c,p})$ have unit vector
$ \hat{{\bm m}}_i^\alpha = (\sin\theta_i^\alpha \cos\varphi^\alpha ,\sin\theta_i^\alpha\sin\varphi^\alpha ,\cos\theta_i^\alpha)$. The $n$ dipoles are located at the vertices of a zig-zag lattice made out of two sublattices that are coplanar parallel chains: chain \emph{c} with $n_\emph{c}$ dipoles and chain \emph{p} with $n_\emph{p}=n_\emph{c}-1$ as shown in Fig.\ref{f1}. Dipoles rotate in an easy plane in terms of a polar angle $\theta$ with respect to the $\hat{z}$ axis, and a fixed azimuthal angle $\varphi^\alpha: \varphi^\emph{c}=0$ and $\varphi^\emph{p}=\frac{\pi}{2}$. Hence, dipoles in \emph{c} rotate in the  $\hat{x}-\hat{z}$ plane and dipoles in \emph{p} in the  $\hat{y}-\hat{z}$ plane (Fig.\ref{f1}).
\begin{figure}
\includegraphics[width=\columnwidth]{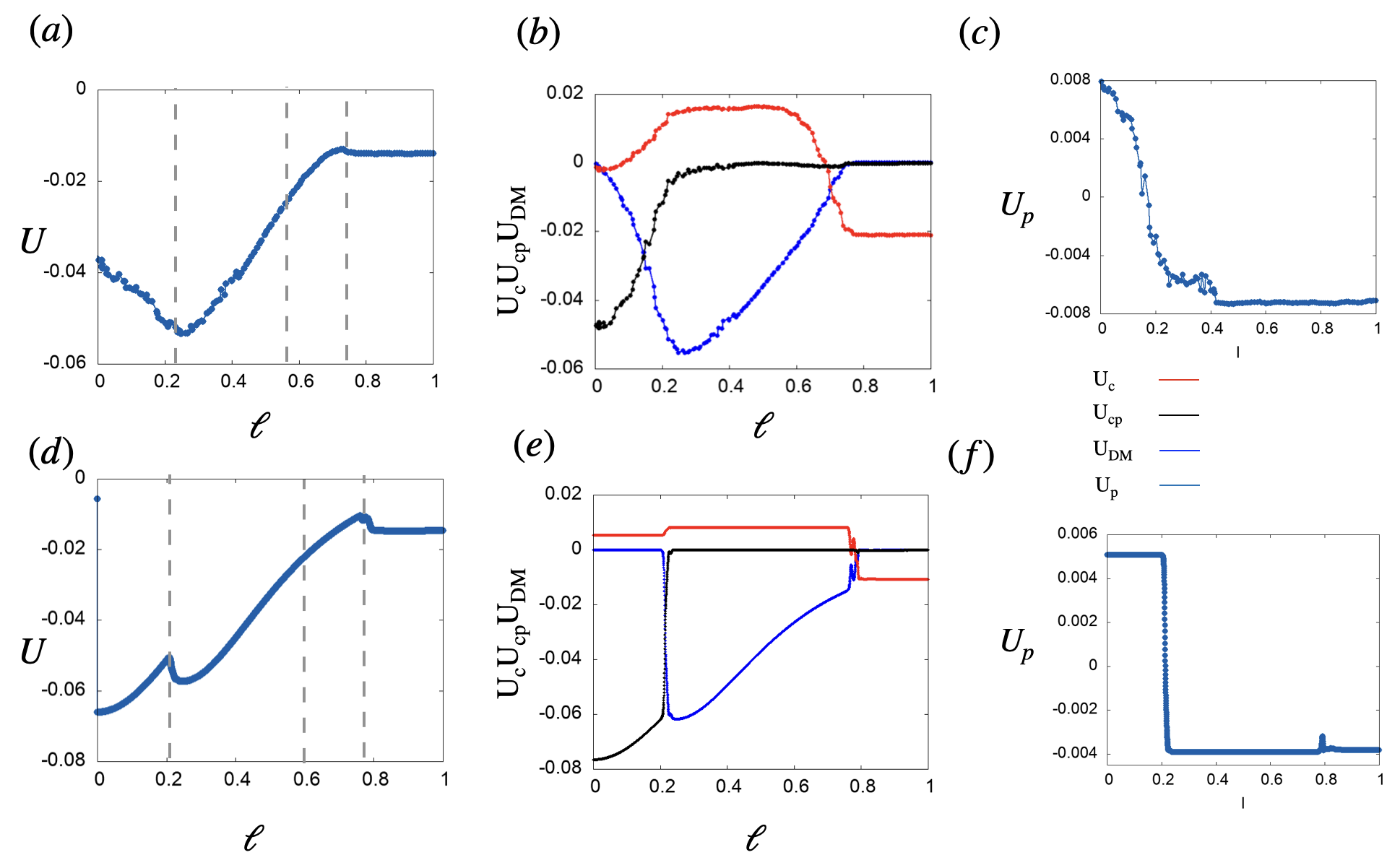}
\caption{(a) and (d) show the total energy of the system versus $\ell$ in experiments and numerics, respectively. The magnetic phases are delimited by the dotted vertical lines. Using Eq.\ref{eq1} the contribution of symmetric and antisymmetric long-range interactions to the total magnetic energy is shown in (b) (experiments) and (e) (numerics) with $U_\emph{c}$,  $U_{\emph{c}\emph{p}}$ and  $U_{DM}$ shown in red, black and blue respectively. The dark blue curves of (c) and (f) show $U_\emph{p}$ in experiments and numerics, respectively.}
\label{f4}
\end{figure}
\begin{figure*}
\includegraphics[width=\textwidth]{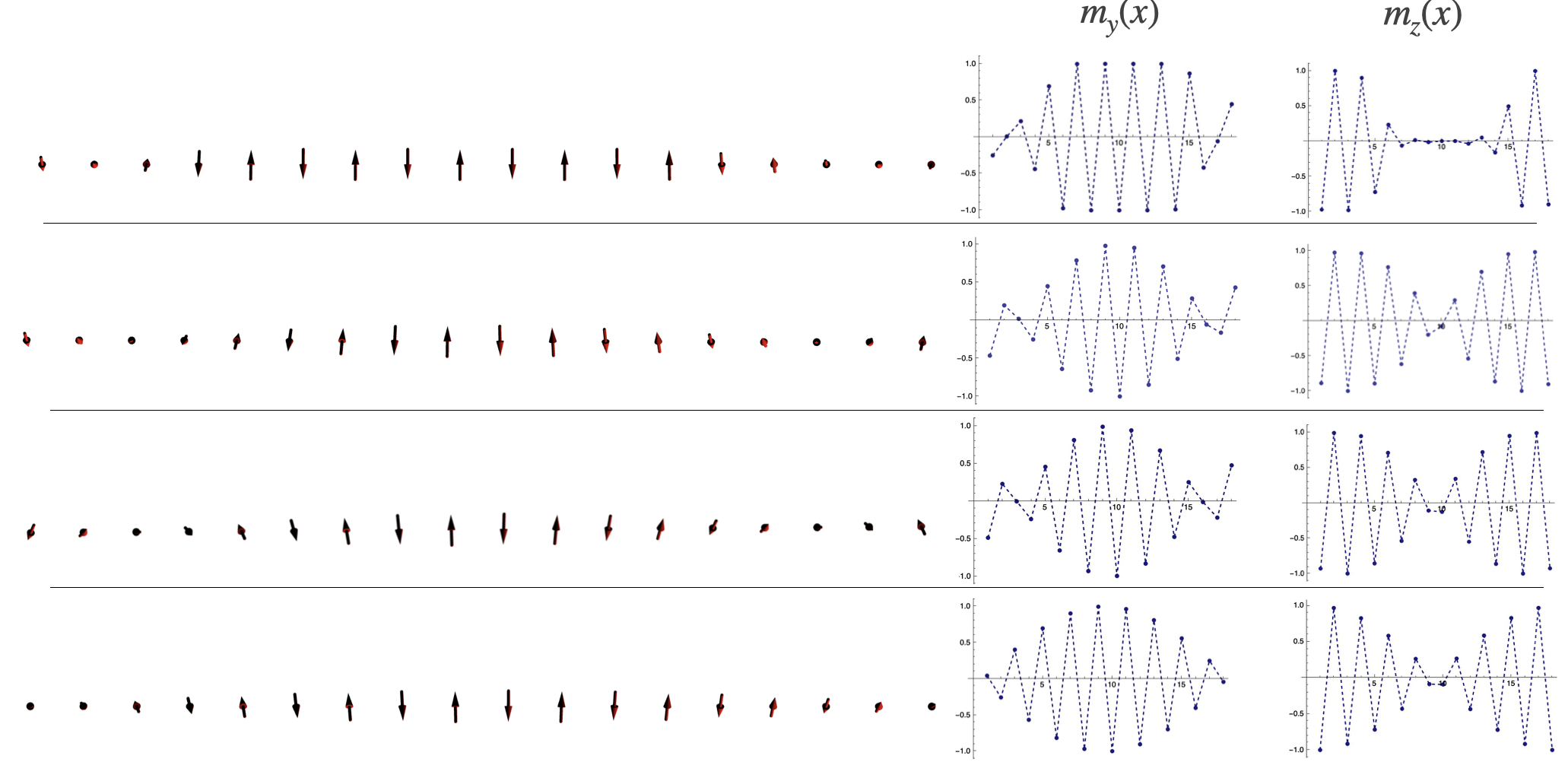}
\caption{Solitons in the \emph{p} sublattice. From top to bottom the left panel shows the winding texture for different values of $\ell$ as it grows from $\ell\sim \ell_*$ up to $\ell\sim0.95$. The right panel shows the magnetization of each dipole in \emph{p} along $\hat{y}$ and  $\hat{z}$ directions.}
\label{f5}
\end{figure*}
With easy planes mutually perpendicular among chains,  the dipolar energy is exactly separable into symmetric and antisymmetric long range interactions:
\begin{widetext}
\begin{align}
U&=&\frac{g}{2}\sum_{i\neq k}^n\left[ J_{ik}^{0}\left(\frac{3}{2}\cos(\theta_i^\emph{c} +\theta_k^\emph{c})-\frac{1}{2}\hat{m}_i^\emph{c}\cdot\hat{m}_k^\emph{c}\right) +J_{ik}^{0}\left(\hat{m}_i^\emph{p}\cdot\hat{m}_k^\emph{p}\right)+J_{ik}\left(\hat{m}_i^\emph{c}\cdot\hat{m}_k^\emph{p}\right) +\bm{\mathcal{D}}_{ik}\cdot(\hat{m}_i^\emph{c}\times \hat{m}_k^\emph{p})\right]
\label{eq1}
\end{align}
\end{widetext}
which give rise to  four energetic contributions to the full magnetic energy of the system are consecutively denoted such that $\rm U=U_\emph{c}+U_\emph{p}+U_{\emph{c}\emph{p}}+U_{DM}$. They correspond respectively to symmetric intra-sublattice interactions  in \emph{c} and  \emph{p}, a symmetric inter-sublattice interaction, and an antisymmetric inter-sublattice interaction energy. Explicit formulas for the associated couplings read, $J_{ik}^{0}=\frac{1}{|i-k|^3}$, $J_{ik}=\frac{1}{\left(\ell^2+(i-k+\frac{1}{2})^2\right)^{3/2}}$ which are respectively symmetric intra-chain and interchain couplings. $\bm{\mathcal{D}}_{ik}=-3\left(0,0,\frac{\ell(i-k+\frac{1}{2})}{\left((i-k+\frac{1}{2})^2+\ell^2\right)^{\frac{5}{2}}}\right)$ corresponds to an interchain Dzyaloshinskii–Moriya antisymmetric coupling, perpendicular to the plane of the lattice. Note the dependence of $J_{ik}$ and $\bm{\mathcal{D}}_{ik}$ on $\ell$.

\section{\label{sec:phases}Magnetic phases in terms of $\ell$}
The experimental setup comprises $n=37$  Neodymium cylindrical magnets of length $a$, radius $r$ and $m_0= a r^2 \pi M_s $, hinged at the sites of a Polytetrafluoroethylene (PTFE) plate forming a zig-zag lattice with lattice constant $\Delta$. Sublattices \emph{c} and \emph{p} have respectively $n_\emph{c}=19$ and $n_\emph{p}=18$ magnets (Figs.\ref{f1}) and rotate in the mutually perpendicular  planes  $x-z$ and  $y-z$ respectively.  A small amount of disorder due to deviations of the dipoles with respect to their easy plane of rotation follows a Gaussian distribution centered at zero and with a standard deviation $\delta\phi\sim 0.005$. The interchain gap $\ell$ is tuned by moving chain  \emph{p} along the $\hat{y}$  axis in the range $\ell\in (0,1.5)$ at a constant speed $v$ while \emph{c} remains at rest. A camera captures magnetic configurations of the system (see Methods) as the stage with chain \emph{p} is moved from $\ell = 1.5$ to $0$ (approaching) and back to $\ell=1.5$ (receding). In Figs.\ref{f1}(c-e) we indicate the north poles of the magnets with a black tip. 

Depending on $\ell$, dipoles settle into four magnetic configurations as shown in Fig.\ref{f1}. At small gaps $\ell\in (0,0.2)=(0,\ell_{f})$ the system realizes an out of plane antiferromagnetic parallel phase, $\rm AF$, along $\hat{z}$. It consists of dipoles arranged ferromagnetically with respect to the others in the same sublattice  and antiferromagnetically with respect to dipoles in the other sublattice. Increasing $\ell$ triggers a spin flop transition where all dipoles depart from $\hat{z}$ and settle in the $x-y$ plane featuring phase $\rm AF^2$  ($\ell\in (0.2,0.6)=(\ell_{f},\bar{\ell})$), where chain \emph{c} describes a collinear antiferromagnetic state, and chain \emph{p} a parallel antiferromagnetic configuration. For intermediate interchain gaps, $\ell\in (0.6,0.8)=(\bar{\ell},\ell_*)$, phase  $\rm AF^2$ competes with phase $\rm FAF$ which differs from $\rm AF^2$ in that chain \emph{c} arranges in a ferromagnetic collinear fashion. At large gaps,  $\ell_*<\ell<1$  chain \emph{c} settles in a collinear ferromagnetic state, while chain \emph{p} defines a winding texture consisting of a twisted parallel antiferromagnetic state in the $y-z$ plane. This phase is denoted Tw and is shown in Fig.\ref{f1}(a-b).  
The previous magnetic configurations define the magnetization curves shown in Fig.\ref{f2}. At small gap $\ell\in (0,\ell_{f})$ Fig.\ref{f2} shows the absence of average magnetization along $\hat{x}$, $\rm M_x$ (a). Instead, in each sublattice the magnetization along $\hat{z}$ has a different sign and reaches its maximum in this regime (bottom panel of (b) and top panel of (c)). Finite staggered magnetizations along $\hat{x}$,  $\rm N_x$ ((b) top panel) for this size of the gap shows that dipoles in \emph{c} realize a canted state in the $x-z$ plane, consistent with Fig.\ref{f1}. Indeed, the $\rm AF$ magnetic state is difficult to accomplish in experiments due to the strong dipolar interactions among nearest neighbor dipoles in different sublattices, the azimuthal disorder and the frictional rotation of the magnets.
For $\ell\in (\ell_{f},\bar{\ell})$ dipoles are in the $x-y$ plane so that $\rm M_z$ goes to zero in both sublattices and staggered magnetizations $\rm N_{x}$ (and $\rm N_{y}$ not shown) reach their maximum values Fig.\ref{f2}(b).  The metastable regime \cite{supp} with competing phases $\rm AF^{2}$ and $\rm FAF$ originates the hysteresis loops  of Figs.\ref{f2}(a-b) at intermediate $\ell\in (\bar{\ell},\ell_{*})$. Finally at large values of the gap $\ell_*<\ell<1$ the \emph{c} chains remain collinear in the magnetic state which results in $\rm M_x=1$ and the winding texture in \emph{p} is such that $\rm N_{y}\neq 0$ and $\rm N_{z}^{p}\neq 0$ while $\rm M_{z}=0$. We further examined the evolution of the system with $\ell$ by implementing molecular dynamics simulations (Methods). Magnetic phases from numerics coincide with those found in experiments as shown by the screenshots of the numerical lattice in Figs.\ref{f1}(b). The hysteretic behavior seen in experiments at intermediate $\ell$ is captured too by simulations as can be verified in the magnetization loops of $\rm M_x$ and $\rm N_{x}$ shown in supplementary Fig.1. The width of the loops is well reproduced by considering a nearest neighbor interacting model \cite{supp}. In addition to the loop at $\ell\in(\bar{\ell},\ell_*)$, numerics reveals another loop for $\ell\in (\ell_{fn},\ell_{f})$. It shows that phases $\rm AF^2$ and $\rm AF$ are metastable in this regime. This is consistent with the spin flop transition being of the first order type \cite{antropov2021tunable,welp1999direct}. 
 In the limit of a large gap, $\ell>1$ we find that the twisted state relaxes into the $\rm FAF$ phase.

\section{\label{sec:couplings}Symmetric and antisymmetric contributions to the dipolar energy.}
Eq.\ref{eq1} reveals the specific contribution of the symmetric and antisymmetric inter and intra-sublattice interactions to the system's total energy. Fig. \ref{f4} shows the evolution of each of them with $\ell$ in units of $\frac{g}{2}$. In Figs. \ref{f4}(a) (experiment) and (c) (numerics) the minimum of the total dipolar energy U  occurs at the onset of phase $\rm AF^2$. This extreme in U coincides with the optimum of $U_{\rm DM}$ (blue curves in Figs.\ref{f4}(b) and (d))  and with the maximum of the spin current along the $\hat{z}$ direction as shown in the bottom panel of Fig.\ref{f2}(c).  Being proportional to the $\hat{z}$ projection of dipoles in chain \emph{c}, the energy $\rm U_\emph{cp}$ (in black) is non null during phase $\rm AF$ and it is the dominant contribution to U at $\ell<\ell_{f}$. From $\ell>\ell_{f}$ and up to $\ell<\bar{\ell}$,  $\rm U_{\rm DM}$ (in blue) dominates the dynamics followed by the intrachain symmetric energy 
$\rm U_{\emph{c}}$ (red) which becomes the dominant contribution to the energy once the system is in the Tw phase. 
Finally, Figs.\ref{f4}(c) and (f) show the intra-chain symmetric energy in \emph{p}, $\rm U_\emph{p}$ to be non null and one order of magnitude smaller than the others in all the range of $\ell$.
At $\ell>\ell_*$ the system is in the Tw phase and the total energy, dominated by $\rm U_{\emph{c}}$, barely changes with $\ell$. Note that in this magnetic phase the dipoles arrange such that the interchain energy contributions cancel out $\rm U_{\rm DM}=U_\emph{cp}=0$. Experiments (Fig.\ref{f4}a,b,c) and numerics (Fig.\ref{f4}d,e,f) agree fairly well. 
Next, consider the case of a long zig-zag lattice. We denote the horizontal distance between two dipoles $x$ and the vertical gap $y$. The Dzyaloshinskii–Moriya coupling perpendicular to the plane of the system is written as, $\bm{\mathcal{D}}=-3\frac{y(x+\frac{1}{2})}{\left((x+\frac{1}{2})^2+y^2\right)^{\frac{5}{2}}}\hat{z}$ and decays fast with dipole distance (see supplementary Fig.6 \cite{supp}).  $\mathcal{D}$ reaches a maximum  for nearest neighbor dipoles ($x=0$)  at an optimum interchain distance $y_{m}=\frac{1}{4}(1 + 2 x)=\frac{1}{4}$, and its contribution to the total energy is comparable with that of the symmetric energies \cite{supp}. Integrating out the x coordinate yields an explicit  formula for the effective gap-dependent interchain chiral coupling in the system, $\mathcal{D}_{\rm ef}=\frac{8 y}{\left(4 y^2+1\right)^{3/2}}$. Similar to the previous case, the symmetric inter and intrachain couplings, $J$ and $J^0$ decay very fast with x and y, but at small $(x,y)$, $\mathcal{D}$ becomes the largest among the three.  The formula for the effective gap dependent interchain symmetric coupling after integrating out the x coordinate yields $J_{\rm ef}=\frac{1-\frac{1}{\sqrt{4 y^2+1}}}{y^2}$ which is maximum at $y=0$. As shown in Fig.1 \cite{supp}, $J(\ell)_{\rm ef}>\mathcal{D}(\ell)_{\rm ef}$ for $\ell<\ell_{nf}$ point at which they are equal and from there $\mathcal{D}(\ell)_{\rm ef}$ becomes the leading coupling.

\section{\label{sec:current}Intrinsic magnetic Current and torque}
 In \cite{katsura2005}, the spin current arises from the Heisenberg equation \cite{katsura2005} whose correspondence to the classical system at hand is:
$\frac{\partial {\bm m}_i}{\partial t} = {\bm m}_i \times  \frac{\partial U}{\partial  {\bm m}_i}= {\bm m}_i \times  {\bm H}_i=\bm{\mathcal{T}}_i= -\sum_k {\bm {\mathcal J}}_{ki}$, where ${\bf H}_i$ denotes the internal magnetic field produced by all dipoles  but the $i$-$th$ at the position of $\hat{\bf m}_{i}$, and $\bm{\mathcal{T}}_i$ is the associated torque. Hence, a magnetic current is induced by the internal magnetic torque, which can be tuned by variating $\ell$. Writing down U in terms of $\mathcal{\bar{I}}_{ik}$,  the interaction matrix of the system, \cite{supp} yields $U=- \frac{1}{2} \sum_{i,k}   \bm{m}_i\mathcal{\bar{I}}_{ik}  \bm{m}_k $, which allows to formulate the classical correspondence of the spin current in terms of $\mathcal{\bar{I}}_{ik}$ as follows, 
\begin{eqnarray}
 {\mathcal J}_{ik}^{(a)} =   \epsilon^{abc}  m_i^{(b)}  \mathcal{\bar{I}}_{ik}^{(c,d)} m_k^{(d)}.
\label{eq2}
\end{eqnarray}
In  Eq.\ref{eq2} `$a$' labels the vector components $x,y,z$ and the matrix elements of $\mathcal{\bar{I}}_{ik}$ correspond to the magnetic interactions that couple dipoles in Eq.\ref{eq1} as shown in \cite{supp}. $\epsilon^{abc} $ denotes the Levi-Civita  symbol, and thus Eq.\ref{eq2} demonstrates that a magnetic current arises from the matrix elements of $\mathcal{\bar{I}}_{ik}$ connecting magnetization vectors of dipoles coupled by  the chiral DM vector $\mathcal{D}$. 

\section{\label{sec:planar}Magnetic Current in the planar states, $\ell_{f}<\ell<\ell_*$}.
Figs.\ref{f2}(c) (experiments) and (f) (numerics) show the $z$ component of the magnetic current in experiments and simulations, $\mathcal {J}^{(z)}=2\sum_{i\in \rm{\emph c}}\sum_{k\in \emph p}\mathcal{D}_{ik}m_i^{x}m_k^{y}$. $\mathcal {J}^{(z)}$ connects magnets whose magnetization have perpendicular components in the $\hat{x}-\hat{y}$ plane. It is zero or negligible at $\ell<\ell_{f}$, and after reaching its maximum in phase $\rm AF^{2}$ it becomes zero once again in phase Tw. As expected from the magnetization loops, $\mathcal {J}^{(z)}$ realizes hysteresis in the metastable regime $(\bar{\ell}<\ell<\ell_*)$.

\subsection{$\mathcal{D}$ as a vector potential to the magnetic current} 
Using Eq.\ref{eq2}, the magnetic current in phase $\rm AF^{2}$ yields, 
\begin{eqnarray} 
\bm{{\mathcal J}}_{\rm AF^{2}}=2g\left[0,0,\sum_{i\neq k, \alpha\neq\beta}\mathcal{D}_{ik}\sin({\theta_i^{\alpha}-\theta_k^{\beta}+\frac{\pi}{2}})\right]
\label{eq3}
\end{eqnarray} 
where $2g\bm{\mathcal{D}}_{i,k}$ is interpreted as the spin stiffness or magnetic rigidity of the system  \cite{katsura2005}.
In  phase $\rm AF^{2}$ the energy of the system $\rm U_{\rm AF^{2}}=U_{\emph{c}}+U_{\emph{p}}+U_{\rm DM}$ (Eq.\ref{eq1})  it yields (in units of $\frac{g}{2}$), (Figs.3(b) and (d)),
\begin{eqnarray}
U_{\rm AF^{2}}&=&
\sum_{i\neq k,(\alpha, \beta=\emph{c,p})}\left[\delta^{\alpha\beta}J_{ik}^{\alpha\beta}+(1-\delta^{\alpha\beta})\frac{\mathcal{D}_{ik}}{2}\right]\sin{\theta_i^{\alpha}}\sin{\theta_k^{\beta}} \nonumber
\end{eqnarray}
where $J_{ik}^{\emph{cc}}=-2J_{ik}^{0}$ and $J_{ik}^{\emph{pp}}=J_{ik}^{0}$. It can be written as $U_{\rm AF^{2}}=\sum(\chi_{ik}^{\alpha\beta}\cos{A_{ik}})\sin{\theta_i^{\alpha}}\sin{\theta_k^{\beta}}$  \cite{supp}, where $\chi_{ik}^{\alpha\beta}\cos{A_{ik}}=\left[\delta^{\alpha\beta}J_{ik}^{\alpha\beta}+(1-\delta^{\alpha\beta})\frac{\mathcal{D}_{i,k}}{2}\right]$. Correspondingly the magnetic current becomes ${\mathcal J}_{\rm AF^{2}}=2\sum\chi_{ik}^{\alpha\beta}\sin{(\theta_i^{\alpha}-\theta_k^{\beta}+\frac{\pi}{2}+A_{ik})}$. Consequently, the chiral DM vector $\bm{\mathcal{D}}$ acts as the vector potential or gauge field associated to the magnetic current,   \cite{katsura2005}. Further,  $\bm{\mathcal{D}}$ gives rise to the magnetic field along the $\hat{x}$ axis, $\bm{H}_{i}^{\rm DM}=\sum_{k}\bm{m}_k\times\bm{\mathcal{D}}_{ik}$ which acts on dipole $i$ in the \emph{c} sublattice. In a large zig-zag lattice $\bm{H}^{\rm DM}$ acts as an effective interchain field $\bm{H}_{\rm ef}^{\rm DM}={\bf N_y}\times\bm{\mathcal{D}}$ whose magnitude is proportional to the staggered magnetization along $\hat{y}$ in chain \emph{p} and to the chiral DM coupling. This effective field $\bm{H}_{\rm ef}^{\rm DM}$  originates a magnetic flux $\Phi^{\rm DM}$ in a loop S parallel to the $y-z$ plane. Because \emph{p} moves at a velocity $v=\frac{dy}{dt}$ with respect to \emph{c}, the time derivative of $\Phi^{\rm DM}$  induces a fem and an electric field $\bm{E}^{\rm DM}$ in such a loop due to Faraday`s law:  $\mathcal{E}^{\rm DM}=\bm{v}\times \bm{H}_{\rm ef}^{\rm DM}=\oint\bm{E}^{\rm DM}\cdot \bm{dS}$. The induced electric field points along the $\hat{z}$ direction and is proportional to the relative speed between sublattices and the magnitude of the DM vector, $\bm{E}^{\rm DM}=\frac{v \mathcal{D}N_{y}}{2\pi}=\frac{v H_{\rm ef}^{\rm DM}}{2\pi}$. This leads to the coupling between the magnetic current and the induced electric field as a gauge field through the electric polarization: $\mathcal{\bf P}=\frac{\partial U_{\rm AF^2}}{\partial{\bm E}^{DM}}$, showing a route for the realization of the magnetoelectric effect in this system.  
\section{\label{sec:solo}Chiral soliton at $\ell>\ell_*$} 
The nonlinearity of the spin dynamics of magnets is primarily determined by the purely geometric properties of the magnetization field that one sublattice exerts on the other. These properties give rise to topologically non-trivial structures in the zig-zag chain of dipoles.  
At $\ell_*<\ell<1$, dipoles belonging to different sublattices remain orthogonal with respect to each other, and thus the equilibrium orientation of the dipoles  minimizes  $U_{\rm AF^{2}}$. Correspondingly, in this regime, the magnetic current $\propto \sin{(\theta_i^{\alpha}-\theta_k^{\beta}+\frac{\pi}{2}+A_{ik})}$ remains zero.  
Once sublattice \emph{c} settles into the ferromagnetic collinear state at $\ell_*$,  the DM energy can be rewritten via the internal effective Dzyaloshinskii field, felt by dipoles in \emph{p} due to their antisymmetric interaction with collinear ferromagnetic dipoles in \emph{c}: $\mathcal{D}(x,\ell)M_x\hat{y}$. When the two sublattices are not farther than $\Delta$, this field destroys the parallel antiferromagnetic state of \emph{p}  because the Zeeman energy orient dipoles along the Dzyaloshinskii field produced by \emph{c}. Further, the effective Dzyaloshinskii field  acts as an anisotropy internal field that rotates dipoles in \emph{p}  along the $\hat{x}$ axis, giving rise to a transverse magnetization along $\hat{z}$.  Consider the Neel order in \emph{p} constrained in the $y-z$ plane, $N_p=(0,\sin\theta,\cos\theta)$. The angle  $\theta(x,t)$ parametrizes the local magnetic state where again $t=\ell/v$. As shown in Fig.\ref{f4}, in phase Tw the full energy remains constant while $U_{DM}=U_{cp}=0$. Therefore $\theta(x,t)$ is such that minimizes $U_c+U_p$ subject to the constraint $U_{DM}=U_{cp}=0$. Because of the fast decay of the couplings $\mathcal{D}$ and $J$ \cite{supp} we consider interchain interactions up to nearest neighbors dipoles. Denoting $\omega_1=(J^{0}(1)+J(1,\ell)+\mathcal{D}(1,\ell))$ and $\omega_2=(J(2,\ell)+\mathcal{D}(2,\ell)+\sum_2^nJ^{0}(x))$ (see \cite{supp} for details), the magnetic texture is given by the solution to the equation $\omega_1 \sin (\theta)+\omega_2 \sin (2 \theta)=0$ which yields $\theta(x,t)=2\arctan{\pm\frac{\sqrt{4\omega_2^{2}-\omega_1^{2}}}{\omega_1}}$, that corresponds to a one dimensional soliton \cite{galkina2018dynamic,hone1980spin,cuevas2014sine}.
The evolution of the  twisted structure as  $\ell$ grows from $\ell_*$ up to 1 and the values of $m_z$ and $m_y$ at each position of \emph{p} are shown in Fig.\ref{f5}.  The soliton consists of two domain walls (each Bloch domain wall is realized by one or more dipoles that have rotated toward the $\hat{z}$ axis) which are born near the edges of \emph{p} at $\ell\sim\ell_*$. The net total topological charge is a conserved quantity, and the associated continuity equation \cite{galkina2018dynamic} defines the dynamics of the winding texture as $\ell$ is tuned, as shown in Fig.\ref{f5}. As $\ell$ grows from $\ell_*$, the two domain walls extend toward the center of \emph{p}, including more dipoles, until they merge. Once they merge at $\ell\sim 1$, the domain walls disappear, and the sublattice is such that all the dipoles orient along the same axis. These winding structures are the product of the internal chiral field and have associated a handedness determined by the sense of $M_x$ in \emph{c}. 
 
 \section{\label{sec:conclusion}Conclusions} 
We have shown that the dynamics of a  zig-zag lattice of dipoles is induced by a magnetic torque, which arises due to an intrinsic DMI between its sublattices without the aid of external sources to break time-reversal symmetry.  The hysteretic dynamics in the system is propelled  by interlayer gap variations that tune the internal chiral and achiral fields. The Dzyaloshinskii-Moriya interaction acts as the vector potential of the magnetic current perpendicular to the plane of the lattice, inducing magnetic and electric fields, which allows the manifestation of magnetoelectricity and opens the door for the study of the gauge theory in this system. Atomic-scale or mesoscopic spin textures with all broken mirror symmetries and preserved time-reversal symmetry like the twisted magnetic order shown here provide a promising platform to study cross-coupled ferroic orders, magnetic optical activities, and topological transport properties. 

\begin{acknowledgments} P.M. acknowledges support from Fondecyt under Grant No. 1210083. K.H. acknowledges support from the Leading House for the Latin American Region (CLS-HSG) Mobility Grant 2020 and the Swiss National Science Foundation (Projects No. 200020 172774).
\end{acknowledgments}

\appendix
\section{Experiments}
The experimental setup consists of one fixed stage and a computer-controlled movable stage, machined from acrylic plexiglass and covered with Teflon to reduce rotational friction. A camera (Nikon D750) records the rotation of the magnets that are free to move in their respective planes due to carbon fixtures. The NdFeB magnets (grade N42) have a Nickel coating, a diameter $d = 1.59 \pm 0.01 \times 10^{-2}$ m, a length $2a = 12.7 \pm 0.01 \times 10^{-3}$ m, and a mass $M = 0.189 \times 10^{-3}$ kg. The speed of the movable stage was set to 0.1 $mm/s$, while the camera records $59$ frames per second. These images are converted to rotation angles using standard imaging techniques.

To compute the damping coefficient $\eta$, we isolated a single rotor, impulsively applied torque to it, and then recorded its relaxation dynamics. We extracted the evolution of $\theta(t)$ using standard imaging techniques corresponding to damped dynamics without external forcing. The damping is computed directly by fitting it to the solution $\theta \sim \exp{(-t/\tau_D)}$, and thus we estimated the damping time of a single rod to be $\tau_D = 0.83\pm0.18$ s.

\section{Molecular dynamics simulations}
 The equation of motion for the polar angle of the inertial magnet located at site $i$ in sublattice $\alpha$, interacting through the full long-range dipolar potential with all other dipoles in the system, reads:
$
I\frac{d^2 \theta_{i}^\alpha}{dt^2}=\mathcal{T}_i^\alpha -\eta \frac{d \theta_{i}^\alpha}{dt}
$
 where $I$ denotes the moment of inertia of the magnets, $\eta$ is the damping for the rotation of dipoles in the lattice \cite{mellado2012macroscopic} and the time $t=y/v=\ell/v$ with v the constant speed of chain \emph{p}. The first term at the right hand side of the previous equation is the intrinsic magnetic torque due to the action of the internal magnetic field due to the dipolar interaction between all dipoles in the system $\mathcal{T}_i^\alpha=\hat{\bf m}_{i}^\alpha\times \rm{\bf H}_i^\alpha$, where $\rm {\bf H}_i^\alpha=\frac{\partial U}{\partial \hat{\bf m}_i^\alpha}$ denotes the internal magnetic field produced by all dipoles  but the $i$-$th$ at the position of $\hat{\bf m}_{i}^\alpha$.  To solve the previous system of equations, we used a Verlet method with an integration time step $\Delta t=4\times 10^{-6}$ s. This is equivalent to $\sim 5\times10^5$ time steps in each of which $\ell$ changes by $\Delta \ell=5.3\times 10^{-8}$. During the simulation interval for the dynamics where sublattice \emph{p} recedes, the gap increased from $\ell =0$ up to $\ell = \ell_{max}=1.2$. The initial angular positions for the dipoles at $\ell =0$ where $\theta^\emph{c}=0$ and $\theta^\emph{p}=\pi$ with a small amount of random disorder added to allow the dynamics and according to experiments.  We performed another set of simulations to examine the dynamics when \emph{p} approaches \emph{c} with the difference that $\ell$ now is decreased from $\ell_{max}$ back to zero. The system's initial conditions correspond to the final magnetic state of the receding process—the total simulation time corresponded to $2$ s in each case. Experimentally measured parameters for lattice constant, damping, inertia, and magnetic charge were used in all simulations.
The results from the molecular simulations were compared with the energy minimization of the system in terms of $\ell$. The numerical minimization of the total energy of the lattice used the `RandomSearch` method in the numerical minimization routine of Wolfram Mathematica 12.0.
To implement the random disorder for the study of avalanches distribution, we added to the previous equation of motion a random torque which in each case followed a gaussian distribution with standard deviation R details can be found in \cite{supp}. 

\begin{thebibliography}{54}%
\makeatletter
\providecommand \@ifxundefined [1]{%
 \@ifx{#1\undefined}
}%
\providecommand \@ifnum [1]{%
 \ifnum #1\expandafter \@firstoftwo
 \else \expandafter \@secondoftwo
 \fi
}%
\providecommand \@ifx [1]{%
 \ifx #1\expandafter \@firstoftwo
 \else \expandafter \@secondoftwo
 \fi
}%
\providecommand \natexlab [1]{#1}%
\providecommand \enquote  [1]{``#1''}%
\providecommand \bibnamefont  [1]{#1}%
\providecommand \bibfnamefont [1]{#1}%
\providecommand \citenamefont [1]{#1}%
\providecommand \href@noop [0]{\@secondoftwo}%
\providecommand \href [0]{\begingroup \@sanitize@url \@href}%
\providecommand \@href[1]{\@@startlink{#1}\@@href}%
\providecommand \@@href[1]{\endgroup#1\@@endlink}%
\providecommand \@sanitize@url [0]{\catcode `\\12\catcode `\$12\catcode
  `\&12\catcode `\#12\catcode `\^12\catcode `\_12\catcode `\%12\relax}%
\providecommand \@@startlink[1]{}%
\providecommand \@@endlink[0]{}%
\providecommand \url  [0]{\begingroup\@sanitize@url \@url }%
\providecommand \@url [1]{\endgroup\@href {#1}{\urlprefix }}%
\providecommand \urlprefix  [0]{URL }%
\providecommand \Eprint [0]{\href }%
\providecommand \doibase [0]{https://doi.org/}%
\providecommand \selectlanguage [0]{\@gobble}%
\providecommand \bibinfo  [0]{\@secondoftwo}%
\providecommand \bibfield  [0]{\@secondoftwo}%
\providecommand \translation [1]{[#1]}%
\providecommand \BibitemOpen [0]{}%
\providecommand \bibitemStop [0]{}%
\providecommand \bibitemNoStop [0]{.\EOS\space}%
\providecommand \EOS [0]{\spacefactor3000\relax}%
\providecommand \BibitemShut  [1]{\csname bibitem#1\endcsname}%
\let\auto@bib@innerbib\@empty
\bibitem [{\citenamefont {Jiang}\ \emph {et~al.}(2017)\citenamefont {Jiang},
  \citenamefont {Liu}, \citenamefont {Wang}, \citenamefont {Wang},\ and\
  \citenamefont {Jiang}}]{jiang2017fluorescent}%
  \BibitemOpen
  \bibfield  {author} {\bibinfo {author} {\bibfnamefont {Y.}~\bibnamefont
  {Jiang}}, \bibinfo {author} {\bibfnamefont {C.}~\bibnamefont {Liu}}, \bibinfo
  {author} {\bibfnamefont {X.}~\bibnamefont {Wang}}, \bibinfo {author}
  {\bibfnamefont {T.}~\bibnamefont {Wang}},\ and\ \bibinfo {author}
  {\bibfnamefont {J.}~\bibnamefont {Jiang}},\ }\bibfield  {title} {\bibinfo
  {title} {Fluorescent phthalocyanine assembly distinguishes chiral isomers of
  different types of amino acids and sugars},\ }\href@noop {} {\bibfield
  {journal} {\bibinfo  {journal} {Langmuir}\ }\textbf {\bibinfo {volume}
  {33}},\ \bibinfo {pages} {7239} (\bibinfo {year} {2017})}\BibitemShut
  {NoStop}%
\bibitem [{\citenamefont {Takane}\ \emph {et~al.}(2019)\citenamefont {Takane},
  \citenamefont {Wang}, \citenamefont {Souma}, \citenamefont {Nakayama},
  \citenamefont {Nakamura}, \citenamefont {Oinuma}, \citenamefont {Nakata},
  \citenamefont {Iwasawa}, \citenamefont {Cacho}, \citenamefont {Kim} \emph
  {et~al.}}]{takane2019observation}%
  \BibitemOpen
  \bibfield  {author} {\bibinfo {author} {\bibfnamefont {D.}~\bibnamefont
  {Takane}}, \bibinfo {author} {\bibfnamefont {Z.}~\bibnamefont {Wang}},
  \bibinfo {author} {\bibfnamefont {S.}~\bibnamefont {Souma}}, \bibinfo
  {author} {\bibfnamefont {K.}~\bibnamefont {Nakayama}}, \bibinfo {author}
  {\bibfnamefont {T.}~\bibnamefont {Nakamura}}, \bibinfo {author}
  {\bibfnamefont {H.}~\bibnamefont {Oinuma}}, \bibinfo {author} {\bibfnamefont
  {Y.}~\bibnamefont {Nakata}}, \bibinfo {author} {\bibfnamefont
  {H.}~\bibnamefont {Iwasawa}}, \bibinfo {author} {\bibfnamefont
  {C.}~\bibnamefont {Cacho}}, \bibinfo {author} {\bibfnamefont
  {T.}~\bibnamefont {Kim}}, \emph {et~al.},\ }\bibfield  {title} {\bibinfo
  {title} {Observation of chiral fermions with a large topological charge and
  associated fermi-arc surface states in cosi},\ }\href@noop {} {\bibfield
  {journal} {\bibinfo  {journal} {Physical review letters}\ }\textbf {\bibinfo
  {volume} {122}},\ \bibinfo {pages} {076402} (\bibinfo {year}
  {2019})}\BibitemShut {NoStop}%
\bibitem [{\citenamefont {Hasan}\ \emph {et~al.}(2021)\citenamefont {Hasan},
  \citenamefont {Chang}, \citenamefont {Belopolski}, \citenamefont {Bian},
  \citenamefont {Xu},\ and\ \citenamefont {Yin}}]{hasan2021weyl}%
  \BibitemOpen
  \bibfield  {author} {\bibinfo {author} {\bibfnamefont {M.~Z.}\ \bibnamefont
  {Hasan}}, \bibinfo {author} {\bibfnamefont {G.}~\bibnamefont {Chang}},
  \bibinfo {author} {\bibfnamefont {I.}~\bibnamefont {Belopolski}}, \bibinfo
  {author} {\bibfnamefont {G.}~\bibnamefont {Bian}}, \bibinfo {author}
  {\bibfnamefont {S.-Y.}\ \bibnamefont {Xu}},\ and\ \bibinfo {author}
  {\bibfnamefont {J.-X.}\ \bibnamefont {Yin}},\ }\bibfield  {title} {\bibinfo
  {title} {Weyl, dirac and high-fold chiral fermions in topological quantum
  matter},\ }\href@noop {} {\bibfield  {journal} {\bibinfo  {journal} {Nature
  Reviews Materials}\ }\textbf {\bibinfo {volume} {6}},\ \bibinfo {pages} {784}
  (\bibinfo {year} {2021})}\BibitemShut {NoStop}%
\bibitem [{\citenamefont {Shekhar}(2018)}]{shekhar2018chirality}%
  \BibitemOpen
  \bibfield  {author} {\bibinfo {author} {\bibfnamefont {C.}~\bibnamefont
  {Shekhar}},\ }\bibfield  {title} {\bibinfo {title} {Chirality meets
  topology},\ }\href@noop {} {\bibfield  {journal} {\bibinfo  {journal} {Nature
  Materials}\ }\textbf {\bibinfo {volume} {17}},\ \bibinfo {pages} {953}
  (\bibinfo {year} {2018})}\BibitemShut {NoStop}%
\bibitem [{\citenamefont {Monteiro}\ \emph {et~al.}(2015)\citenamefont
  {Monteiro}, \citenamefont {Abanov},\ and\ \citenamefont
  {Kharzeev}}]{monteiro2015magnetotransport}%
  \BibitemOpen
  \bibfield  {author} {\bibinfo {author} {\bibfnamefont {G.~M.}\ \bibnamefont
  {Monteiro}}, \bibinfo {author} {\bibfnamefont {A.~G.}\ \bibnamefont
  {Abanov}},\ and\ \bibinfo {author} {\bibfnamefont {D.~E.}\ \bibnamefont
  {Kharzeev}},\ }\bibfield  {title} {\bibinfo {title} {Magnetotransport in
  dirac metals: Chiral magnetic effect and quantum oscillations},\ }\href@noop
  {} {\bibfield  {journal} {\bibinfo  {journal} {Physical Review B}\ }\textbf
  {\bibinfo {volume} {92}},\ \bibinfo {pages} {165109} (\bibinfo {year}
  {2015})}\BibitemShut {NoStop}%
\bibitem [{\citenamefont {Li}\ \emph {et~al.}(2016)\citenamefont {Li},
  \citenamefont {Kharzeev}, \citenamefont {Zhang}, \citenamefont {Huang},
  \citenamefont {Pletikosi{\'c}}, \citenamefont {Fedorov}, \citenamefont
  {Zhong}, \citenamefont {Schneeloch}, \citenamefont {Gu},\ and\ \citenamefont
  {Valla}}]{li2016chiral}%
  \BibitemOpen
  \bibfield  {author} {\bibinfo {author} {\bibfnamefont {Q.}~\bibnamefont
  {Li}}, \bibinfo {author} {\bibfnamefont {D.~E.}\ \bibnamefont {Kharzeev}},
  \bibinfo {author} {\bibfnamefont {C.}~\bibnamefont {Zhang}}, \bibinfo
  {author} {\bibfnamefont {Y.}~\bibnamefont {Huang}}, \bibinfo {author}
  {\bibfnamefont {I.}~\bibnamefont {Pletikosi{\'c}}}, \bibinfo {author}
  {\bibfnamefont {A.}~\bibnamefont {Fedorov}}, \bibinfo {author} {\bibfnamefont
  {R.}~\bibnamefont {Zhong}}, \bibinfo {author} {\bibfnamefont
  {J.}~\bibnamefont {Schneeloch}}, \bibinfo {author} {\bibfnamefont
  {G.}~\bibnamefont {Gu}},\ and\ \bibinfo {author} {\bibfnamefont
  {T.}~\bibnamefont {Valla}},\ }\bibfield  {title} {\bibinfo {title} {Chiral
  magnetic effect in zrte5},\ }\href@noop {} {\bibfield  {journal} {\bibinfo
  {journal} {Nature Physics}\ }\textbf {\bibinfo {volume} {12}},\ \bibinfo
  {pages} {550} (\bibinfo {year} {2016})}\BibitemShut {NoStop}%
\bibitem [{\citenamefont {Cortijo}\ \emph {et~al.}(2016)\citenamefont
  {Cortijo}, \citenamefont {Kharzeev}, \citenamefont {Landsteiner},\ and\
  \citenamefont {Vozmediano}}]{cortijo2016strain}%
  \BibitemOpen
  \bibfield  {author} {\bibinfo {author} {\bibfnamefont {A.}~\bibnamefont
  {Cortijo}}, \bibinfo {author} {\bibfnamefont {D.}~\bibnamefont {Kharzeev}},
  \bibinfo {author} {\bibfnamefont {K.}~\bibnamefont {Landsteiner}},\ and\
  \bibinfo {author} {\bibfnamefont {M.~A.}\ \bibnamefont {Vozmediano}},\
  }\bibfield  {title} {\bibinfo {title} {Strain-induced chiral magnetic effect
  in weyl semimetals},\ }\href@noop {} {\bibfield  {journal} {\bibinfo
  {journal} {Physical Review B}\ }\textbf {\bibinfo {volume} {94}},\ \bibinfo
  {pages} {241405} (\bibinfo {year} {2016})}\BibitemShut {NoStop}%
\bibitem [{\citenamefont {Kharzeev}\ and\ \citenamefont
  {Liao}(2021)}]{kharzeev2021chiral}%
  \BibitemOpen
  \bibfield  {author} {\bibinfo {author} {\bibfnamefont {D.~E.}\ \bibnamefont
  {Kharzeev}}\ and\ \bibinfo {author} {\bibfnamefont {J.}~\bibnamefont
  {Liao}},\ }\bibfield  {title} {\bibinfo {title} {Chiral magnetic effect
  reveals the topology of gauge fields in heavy-ion collisions},\ }\href@noop
  {} {\bibfield  {journal} {\bibinfo  {journal} {Nature Reviews Physics}\
  }\textbf {\bibinfo {volume} {3}},\ \bibinfo {pages} {55} (\bibinfo {year}
  {2021})}\BibitemShut {NoStop}%
\bibitem [{\citenamefont {Xia}\ \emph {et~al.}(1997)\citenamefont {Xia},
  \citenamefont {Zhang}, \citenamefont {Lu},\ and\ \citenamefont
  {Zhai}}]{xia1997noncollinear}%
  \BibitemOpen
  \bibfield  {author} {\bibinfo {author} {\bibfnamefont {K.}~\bibnamefont
  {Xia}}, \bibinfo {author} {\bibfnamefont {W.}~\bibnamefont {Zhang}}, \bibinfo
  {author} {\bibfnamefont {M.}~\bibnamefont {Lu}},\ and\ \bibinfo {author}
  {\bibfnamefont {H.}~\bibnamefont {Zhai}},\ }\bibfield  {title} {\bibinfo
  {title} {Noncollinear interlayer exchange coupling caused by interface
  spin-orbit interaction},\ }\href@noop {} {\bibfield  {journal} {\bibinfo
  {journal} {Physical Review B}\ }\textbf {\bibinfo {volume} {55}},\ \bibinfo
  {pages} {12561} (\bibinfo {year} {1997})}\BibitemShut {NoStop}%
\bibitem [{\citenamefont {Meckler}\ \emph {et~al.}(2009)\citenamefont
  {Meckler}, \citenamefont {Mikuszeit}, \citenamefont {Pre{\ss}ler},
  \citenamefont {Vedmedenko}, \citenamefont {Pietzsch},\ and\ \citenamefont
  {Wiesendanger}}]{meckler2009real}%
  \BibitemOpen
  \bibfield  {author} {\bibinfo {author} {\bibfnamefont {S.}~\bibnamefont
  {Meckler}}, \bibinfo {author} {\bibfnamefont {N.}~\bibnamefont {Mikuszeit}},
  \bibinfo {author} {\bibfnamefont {A.}~\bibnamefont {Pre{\ss}ler}}, \bibinfo
  {author} {\bibfnamefont {E.}~\bibnamefont {Vedmedenko}}, \bibinfo {author}
  {\bibfnamefont {O.}~\bibnamefont {Pietzsch}},\ and\ \bibinfo {author}
  {\bibfnamefont {R.}~\bibnamefont {Wiesendanger}},\ }\bibfield  {title}
  {\bibinfo {title} {Real-space observation of a right-rotating inhomogeneous
  cycloidal spin spiral by spin-polarized scanning tunneling microscopy in a
  triple axes vector magnet},\ }\href@noop {} {\bibfield  {journal} {\bibinfo
  {journal} {Physical review letters}\ }\textbf {\bibinfo {volume} {103}},\
  \bibinfo {pages} {157201} (\bibinfo {year} {2009})}\BibitemShut {NoStop}%
\bibitem [{\citenamefont {von Bergmann}\ \emph {et~al.}(2014)\citenamefont {von
  Bergmann}, \citenamefont {Kubetzka}, \citenamefont {Pietzsch},\ and\
  \citenamefont {Wiesendanger}}]{von2014interface}%
  \BibitemOpen
  \bibfield  {author} {\bibinfo {author} {\bibfnamefont {K.}~\bibnamefont {von
  Bergmann}}, \bibinfo {author} {\bibfnamefont {A.}~\bibnamefont {Kubetzka}},
  \bibinfo {author} {\bibfnamefont {O.}~\bibnamefont {Pietzsch}},\ and\
  \bibinfo {author} {\bibfnamefont {R.}~\bibnamefont {Wiesendanger}},\
  }\bibfield  {title} {\bibinfo {title} {Interface-induced chiral domain walls,
  spin spirals and skyrmions revealed by spin-polarized scanning tunneling
  microscopy},\ }\href@noop {} {\bibfield  {journal} {\bibinfo  {journal}
  {Journal of Physics: Condensed Matter}\ }\textbf {\bibinfo {volume} {26}},\
  \bibinfo {pages} {394002} (\bibinfo {year} {2014})}\BibitemShut {NoStop}%
\bibitem [{\citenamefont {Komineas}\ and\ \citenamefont
  {Papanicolaou}(2015)}]{komineas2015skyrmion}%
  \BibitemOpen
  \bibfield  {author} {\bibinfo {author} {\bibfnamefont {S.}~\bibnamefont
  {Komineas}}\ and\ \bibinfo {author} {\bibfnamefont {N.}~\bibnamefont
  {Papanicolaou}},\ }\bibfield  {title} {\bibinfo {title} {Skyrmion dynamics in
  chiral ferromagnets},\ }\href@noop {} {\bibfield  {journal} {\bibinfo
  {journal} {Physical Review B}\ }\textbf {\bibinfo {volume} {92}},\ \bibinfo
  {pages} {064412} (\bibinfo {year} {2015})}\BibitemShut {NoStop}%
\bibitem [{\citenamefont {Yang}\ \emph {et~al.}(2021)\citenamefont {Yang},
  \citenamefont {Naaman}, \citenamefont {Paltiel},\ and\ \citenamefont
  {Parkin}}]{yang2021chiral}%
  \BibitemOpen
  \bibfield  {author} {\bibinfo {author} {\bibfnamefont {S.-H.}\ \bibnamefont
  {Yang}}, \bibinfo {author} {\bibfnamefont {R.}~\bibnamefont {Naaman}},
  \bibinfo {author} {\bibfnamefont {Y.}~\bibnamefont {Paltiel}},\ and\ \bibinfo
  {author} {\bibfnamefont {S.~S.}\ \bibnamefont {Parkin}},\ }\bibfield  {title}
  {\bibinfo {title} {Chiral spintronics},\ }\href@noop {} {\bibfield  {journal}
  {\bibinfo  {journal} {Nature Reviews Physics}\ }\textbf {\bibinfo {volume}
  {3}},\ \bibinfo {pages} {328} (\bibinfo {year} {2021})}\BibitemShut {NoStop}%
\bibitem [{\citenamefont {Liang}\ \emph {et~al.}(2018)\citenamefont {Liang},
  \citenamefont {Lin}, \citenamefont {Kushwaha}, \citenamefont {Xing},
  \citenamefont {Ni}, \citenamefont {Cava},\ and\ \citenamefont
  {Ong}}]{liang2018experimental}%
  \BibitemOpen
  \bibfield  {author} {\bibinfo {author} {\bibfnamefont {S.}~\bibnamefont
  {Liang}}, \bibinfo {author} {\bibfnamefont {J.}~\bibnamefont {Lin}}, \bibinfo
  {author} {\bibfnamefont {S.}~\bibnamefont {Kushwaha}}, \bibinfo {author}
  {\bibfnamefont {J.}~\bibnamefont {Xing}}, \bibinfo {author} {\bibfnamefont
  {N.}~\bibnamefont {Ni}}, \bibinfo {author} {\bibfnamefont {R.~J.}\
  \bibnamefont {Cava}},\ and\ \bibinfo {author} {\bibfnamefont {N.~P.}\
  \bibnamefont {Ong}},\ }\bibfield  {title} {\bibinfo {title} {Experimental
  tests of the chiral anomaly magnetoresistance in the dirac-weyl semimetals na
  3 bi and gdptbi},\ }\href@noop {} {\bibfield  {journal} {\bibinfo  {journal}
  {Physical Review X}\ }\textbf {\bibinfo {volume} {8}},\ \bibinfo {pages}
  {031002} (\bibinfo {year} {2018})}\BibitemShut {NoStop}%
\bibitem [{\citenamefont {Claborn}\ \emph {et~al.}(2008)\citenamefont
  {Claborn}, \citenamefont {Isborn}, \citenamefont {Kaminsky},\ and\
  \citenamefont {Kahr}}]{claborn2008optical}%
  \BibitemOpen
  \bibfield  {author} {\bibinfo {author} {\bibfnamefont {K.}~\bibnamefont
  {Claborn}}, \bibinfo {author} {\bibfnamefont {C.}~\bibnamefont {Isborn}},
  \bibinfo {author} {\bibfnamefont {W.}~\bibnamefont {Kaminsky}},\ and\
  \bibinfo {author} {\bibfnamefont {B.}~\bibnamefont {Kahr}},\ }\bibfield
  {title} {\bibinfo {title} {Optical rotation of achiral compounds},\
  }\href@noop {} {\bibfield  {journal} {\bibinfo  {journal} {Angewandte Chemie
  International Edition}\ }\textbf {\bibinfo {volume} {47}},\ \bibinfo {pages}
  {5706} (\bibinfo {year} {2008})}\BibitemShut {NoStop}%
\bibitem [{\citenamefont {Moriya}(1960{\natexlab{a}})}]{moriya1960new}%
  \BibitemOpen
  \bibfield  {author} {\bibinfo {author} {\bibfnamefont {T.}~\bibnamefont
  {Moriya}},\ }\bibfield  {title} {\bibinfo {title} {New mechanism of
  anisotropic superexchange interaction},\ }\href@noop {} {\bibfield  {journal}
  {\bibinfo  {journal} {Physical Review Letters}\ }\textbf {\bibinfo {volume}
  {4}},\ \bibinfo {pages} {228} (\bibinfo {year}
  {1960}{\natexlab{a}})}\BibitemShut {NoStop}%
\bibitem [{\citenamefont {Ryu}\ \emph {et~al.}(2013)\citenamefont {Ryu},
  \citenamefont {Thomas}, \citenamefont {Yang},\ and\ \citenamefont
  {Parkin}}]{ryu2013chiral}%
  \BibitemOpen
  \bibfield  {author} {\bibinfo {author} {\bibfnamefont {K.-S.}\ \bibnamefont
  {Ryu}}, \bibinfo {author} {\bibfnamefont {L.}~\bibnamefont {Thomas}},
  \bibinfo {author} {\bibfnamefont {S.-H.}\ \bibnamefont {Yang}},\ and\
  \bibinfo {author} {\bibfnamefont {S.}~\bibnamefont {Parkin}},\ }\bibfield
  {title} {\bibinfo {title} {Chiral spin torque at magnetic domain walls},\
  }\href@noop {} {\bibfield  {journal} {\bibinfo  {journal} {Nature
  nanotechnology}\ }\textbf {\bibinfo {volume} {8}},\ \bibinfo {pages} {527}
  (\bibinfo {year} {2013})}\BibitemShut {NoStop}%
\bibitem [{\citenamefont {Dzyaloshinskii}(1964)}]{dzyaloshinskii1964theory}%
  \BibitemOpen
  \bibfield  {author} {\bibinfo {author} {\bibfnamefont {I.}~\bibnamefont
  {Dzyaloshinskii}},\ }\bibfield  {title} {\bibinfo {title} {Theory of
  helicoidal structures in antiferromagnets. i. nonmetals},\ }\href@noop {}
  {\bibfield  {journal} {\bibinfo  {journal} {Sov. Phys. JETP}\ }\textbf
  {\bibinfo {volume} {19}},\ \bibinfo {pages} {960} (\bibinfo {year}
  {1964})}\BibitemShut {NoStop}%
\bibitem [{\citenamefont
  {Dzyaloshinsky}(1958)}]{dzyaloshinsky1958thermodynamic}%
  \BibitemOpen
  \bibfield  {author} {\bibinfo {author} {\bibfnamefont {I.}~\bibnamefont
  {Dzyaloshinsky}},\ }\bibfield  {title} {\bibinfo {title} {A thermodynamic
  theory of “weak” ferromagnetism of antiferromagnetics},\ }\href@noop {}
  {\bibfield  {journal} {\bibinfo  {journal} {Journal of physics and chemistry
  of solids}\ }\textbf {\bibinfo {volume} {4}},\ \bibinfo {pages} {241}
  (\bibinfo {year} {1958})}\BibitemShut {NoStop}%
\bibitem [{\citenamefont {Moriya}(1960{\natexlab{b}})}]{moriya1960anisotropic}%
  \BibitemOpen
  \bibfield  {author} {\bibinfo {author} {\bibfnamefont {T.}~\bibnamefont
  {Moriya}},\ }\bibfield  {title} {\bibinfo {title} {Anisotropic superexchange
  interaction and weak ferromagnetism},\ }\href@noop {} {\bibfield  {journal}
  {\bibinfo  {journal} {Physical review}\ }\textbf {\bibinfo {volume} {120}},\
  \bibinfo {pages} {91} (\bibinfo {year} {1960}{\natexlab{b}})}\BibitemShut
  {NoStop}%
\bibitem [{\citenamefont {Anderson}(1950)}]{anderson1950antiferromagnetism}%
  \BibitemOpen
  \bibfield  {author} {\bibinfo {author} {\bibfnamefont {P.~W.}\ \bibnamefont
  {Anderson}},\ }\bibfield  {title} {\bibinfo {title} {Antiferromagnetism.
  theory of superexchange interaction},\ }\href@noop {} {\bibfield  {journal}
  {\bibinfo  {journal} {Physical Review}\ }\textbf {\bibinfo {volume} {79}},\
  \bibinfo {pages} {350} (\bibinfo {year} {1950})}\BibitemShut {NoStop}%
\bibitem [{\citenamefont {Shekhtman}\ \emph {et~al.}(1992)\citenamefont
  {Shekhtman}, \citenamefont {Entin-Wohlman},\ and\ \citenamefont
  {Aharony}}]{shekhtman1992moriya}%
  \BibitemOpen
  \bibfield  {author} {\bibinfo {author} {\bibfnamefont {L.}~\bibnamefont
  {Shekhtman}}, \bibinfo {author} {\bibfnamefont {O.}~\bibnamefont
  {Entin-Wohlman}},\ and\ \bibinfo {author} {\bibfnamefont {A.}~\bibnamefont
  {Aharony}},\ }\bibfield  {title} {\bibinfo {title} {Moriya’s anisotropic
  superexchange interaction, frustration, and dzyaloshinsky’s weak
  ferromagnetism},\ }\href@noop {} {\bibfield  {journal} {\bibinfo  {journal}
  {Physical review letters}\ }\textbf {\bibinfo {volume} {69}},\ \bibinfo
  {pages} {836} (\bibinfo {year} {1992})}\BibitemShut {NoStop}%
\bibitem [{\citenamefont {Freimuth}\ \emph {et~al.}(2020)\citenamefont
  {Freimuth}, \citenamefont {Bl{\"u}gel},\ and\ \citenamefont
  {Mokrousov}}]{freimuth2020dynamical}%
  \BibitemOpen
  \bibfield  {author} {\bibinfo {author} {\bibfnamefont {F.}~\bibnamefont
  {Freimuth}}, \bibinfo {author} {\bibfnamefont {S.}~\bibnamefont
  {Bl{\"u}gel}},\ and\ \bibinfo {author} {\bibfnamefont {Y.}~\bibnamefont
  {Mokrousov}},\ }\bibfield  {title} {\bibinfo {title} {Dynamical and
  current-induced dzyaloshinskii-moriya interaction: Role for damping,
  gyromagnetism, and current-induced torques in noncollinear magnets},\
  }\href@noop {} {\bibfield  {journal} {\bibinfo  {journal} {Physical Review
  B}\ }\textbf {\bibinfo {volume} {102}},\ \bibinfo {pages} {245411} (\bibinfo
  {year} {2020})}\BibitemShut {NoStop}%
\bibitem [{\citenamefont {Deger}(2020)}]{deger2020strain}%
  \BibitemOpen
  \bibfield  {author} {\bibinfo {author} {\bibfnamefont {C.}~\bibnamefont
  {Deger}},\ }\bibfield  {title} {\bibinfo {title} {Strain-enhanced
  dzyaloshinskii--moriya interaction at co/pt interfaces},\ }\href@noop {}
  {\bibfield  {journal} {\bibinfo  {journal} {Scientific Reports}\ }\textbf
  {\bibinfo {volume} {10}},\ \bibinfo {pages} {1} (\bibinfo {year}
  {2020})}\BibitemShut {NoStop}%
\bibitem [{\citenamefont {Chen}\ \emph {et~al.}(2021)\citenamefont {Chen},
  \citenamefont {Robertson}, \citenamefont {Hoffmann}, \citenamefont {Ophus},
  \citenamefont {Cauduro}, \citenamefont {Conte}, \citenamefont {Ding},
  \citenamefont {Wiesendanger}, \citenamefont {Bl{\"u}gel}, \citenamefont
  {Schmid} \emph {et~al.}}]{chen2021observation}%
  \BibitemOpen
  \bibfield  {author} {\bibinfo {author} {\bibfnamefont {G.}~\bibnamefont
  {Chen}}, \bibinfo {author} {\bibfnamefont {M.}~\bibnamefont {Robertson}},
  \bibinfo {author} {\bibfnamefont {M.}~\bibnamefont {Hoffmann}}, \bibinfo
  {author} {\bibfnamefont {C.}~\bibnamefont {Ophus}}, \bibinfo {author}
  {\bibfnamefont {A.~L.~F.}\ \bibnamefont {Cauduro}}, \bibinfo {author}
  {\bibfnamefont {R.~L.}\ \bibnamefont {Conte}}, \bibinfo {author}
  {\bibfnamefont {H.}~\bibnamefont {Ding}}, \bibinfo {author} {\bibfnamefont
  {R.}~\bibnamefont {Wiesendanger}}, \bibinfo {author} {\bibfnamefont
  {S.}~\bibnamefont {Bl{\"u}gel}}, \bibinfo {author} {\bibfnamefont {A.~K.}\
  \bibnamefont {Schmid}}, \emph {et~al.},\ }\bibfield  {title} {\bibinfo
  {title} {Observation of hydrogen-induced dzyaloshinskii-moriya interaction
  and reversible switching of magnetic chirality},\ }\href@noop {} {\bibfield
  {journal} {\bibinfo  {journal} {Physical Review X}\ }\textbf {\bibinfo
  {volume} {11}},\ \bibinfo {pages} {021015} (\bibinfo {year}
  {2021})}\BibitemShut {NoStop}%
\bibitem [{\citenamefont {Fern{\'a}ndez-Pacheco}\ \emph
  {et~al.}(2019)\citenamefont {Fern{\'a}ndez-Pacheco}, \citenamefont
  {Vedmedenko}, \citenamefont {Ummelen}, \citenamefont {Mansell}, \citenamefont
  {Petit},\ and\ \citenamefont {Cowburn}}]{fernandez2019symmetry}%
  \BibitemOpen
  \bibfield  {author} {\bibinfo {author} {\bibfnamefont {A.}~\bibnamefont
  {Fern{\'a}ndez-Pacheco}}, \bibinfo {author} {\bibfnamefont {E.}~\bibnamefont
  {Vedmedenko}}, \bibinfo {author} {\bibfnamefont {F.}~\bibnamefont {Ummelen}},
  \bibinfo {author} {\bibfnamefont {R.}~\bibnamefont {Mansell}}, \bibinfo
  {author} {\bibfnamefont {D.}~\bibnamefont {Petit}},\ and\ \bibinfo {author}
  {\bibfnamefont {R.~P.}\ \bibnamefont {Cowburn}},\ }\bibfield  {title}
  {\bibinfo {title} {Symmetry-breaking interlayer dzyaloshinskii--moriya
  interactions in synthetic antiferromagnets},\ }\href@noop {} {\bibfield
  {journal} {\bibinfo  {journal} {Nature materials}\ }\textbf {\bibinfo
  {volume} {18}},\ \bibinfo {pages} {679} (\bibinfo {year} {2019})}\BibitemShut
  {NoStop}%
\bibitem [{\citenamefont {Cr{\'e}pieux}\ and\ \citenamefont
  {Lacroix}(1998)}]{crepieux1998dzyaloshinsky}%
  \BibitemOpen
  \bibfield  {author} {\bibinfo {author} {\bibfnamefont {A.}~\bibnamefont
  {Cr{\'e}pieux}}\ and\ \bibinfo {author} {\bibfnamefont {C.}~\bibnamefont
  {Lacroix}},\ }\bibfield  {title} {\bibinfo {title} {Dzyaloshinsky--moriya
  interactions induced by symmetry breaking at a surface},\ }\href@noop {}
  {\bibfield  {journal} {\bibinfo  {journal} {Journal of magnetism and magnetic
  materials}\ }\textbf {\bibinfo {volume} {182}},\ \bibinfo {pages} {341}
  (\bibinfo {year} {1998})}\BibitemShut {NoStop}%
\bibitem [{\citenamefont {Heide}\ \emph {et~al.}(2008)\citenamefont {Heide},
  \citenamefont {Bihlmayer},\ and\ \citenamefont
  {Bl{\"u}gel}}]{heide2008dzyaloshinskii}%
  \BibitemOpen
  \bibfield  {author} {\bibinfo {author} {\bibfnamefont {M.}~\bibnamefont
  {Heide}}, \bibinfo {author} {\bibfnamefont {G.}~\bibnamefont {Bihlmayer}},\
  and\ \bibinfo {author} {\bibfnamefont {S.}~\bibnamefont {Bl{\"u}gel}},\
  }\bibfield  {title} {\bibinfo {title} {Dzyaloshinskii-moriya interaction
  accounting for the orientation of magnetic domains in ultrathin films: Fe/w
  (110)},\ }\href@noop {} {\bibfield  {journal} {\bibinfo  {journal} {Physical
  Review B}\ }\textbf {\bibinfo {volume} {78}},\ \bibinfo {pages} {140403}
  (\bibinfo {year} {2008})}\BibitemShut {NoStop}%
\bibitem [{\citenamefont {Torrejon}\ \emph {et~al.}(2014)\citenamefont
  {Torrejon}, \citenamefont {Kim}, \citenamefont {Sinha}, \citenamefont
  {Mitani}, \citenamefont {Hayashi}, \citenamefont {Yamanouchi},\ and\
  \citenamefont {Ohno}}]{torrejon2014interface}%
  \BibitemOpen
  \bibfield  {author} {\bibinfo {author} {\bibfnamefont {J.}~\bibnamefont
  {Torrejon}}, \bibinfo {author} {\bibfnamefont {J.}~\bibnamefont {Kim}},
  \bibinfo {author} {\bibfnamefont {J.}~\bibnamefont {Sinha}}, \bibinfo
  {author} {\bibfnamefont {S.}~\bibnamefont {Mitani}}, \bibinfo {author}
  {\bibfnamefont {M.}~\bibnamefont {Hayashi}}, \bibinfo {author} {\bibfnamefont
  {M.}~\bibnamefont {Yamanouchi}},\ and\ \bibinfo {author} {\bibfnamefont
  {H.}~\bibnamefont {Ohno}},\ }\bibfield  {title} {\bibinfo {title} {Interface
  control of the magnetic chirality in cofeb/mgo heterostructures with
  heavy-metal underlayers},\ }\href@noop {} {\bibfield  {journal} {\bibinfo
  {journal} {Nature communications}\ }\textbf {\bibinfo {volume} {5}},\
  \bibinfo {pages} {1} (\bibinfo {year} {2014})}\BibitemShut {NoStop}%
\bibitem [{\citenamefont {Belmeguenai}\ \emph {et~al.}(2015)\citenamefont
  {Belmeguenai}, \citenamefont {Adam}, \citenamefont {Roussign{\'e}},
  \citenamefont {Eimer}, \citenamefont {Devolder}, \citenamefont {Kim},
  \citenamefont {Cherif}, \citenamefont {Stashkevich},\ and\ \citenamefont
  {Thiaville}}]{belmeguenai2015interfacial}%
  \BibitemOpen
  \bibfield  {author} {\bibinfo {author} {\bibfnamefont {M.}~\bibnamefont
  {Belmeguenai}}, \bibinfo {author} {\bibfnamefont {J.-P.}\ \bibnamefont
  {Adam}}, \bibinfo {author} {\bibfnamefont {Y.}~\bibnamefont {Roussign{\'e}}},
  \bibinfo {author} {\bibfnamefont {S.}~\bibnamefont {Eimer}}, \bibinfo
  {author} {\bibfnamefont {T.}~\bibnamefont {Devolder}}, \bibinfo {author}
  {\bibfnamefont {J.-V.}\ \bibnamefont {Kim}}, \bibinfo {author} {\bibfnamefont
  {S.~M.}\ \bibnamefont {Cherif}}, \bibinfo {author} {\bibfnamefont
  {A.}~\bibnamefont {Stashkevich}},\ and\ \bibinfo {author} {\bibfnamefont
  {A.}~\bibnamefont {Thiaville}},\ }\bibfield  {title} {\bibinfo {title}
  {Interfacial dzyaloshinskii-moriya interaction in perpendicularly magnetized
  pt/co/alo x ultrathin films measured by brillouin light spectroscopy},\
  }\href@noop {} {\bibfield  {journal} {\bibinfo  {journal} {Physical Review
  B}\ }\textbf {\bibinfo {volume} {91}},\ \bibinfo {pages} {180405} (\bibinfo
  {year} {2015})}\BibitemShut {NoStop}%
\bibitem [{\citenamefont {Gross}\ \emph {et~al.}(2016)\citenamefont {Gross},
  \citenamefont {Mart{\'\i}nez}, \citenamefont {Tetienne}, \citenamefont
  {Hingant}, \citenamefont {Roch}, \citenamefont {Garcia}, \citenamefont
  {Soucaille}, \citenamefont {Adam}, \citenamefont {Kim}, \citenamefont
  {Rohart} \emph {et~al.}}]{gross2016direct}%
  \BibitemOpen
  \bibfield  {author} {\bibinfo {author} {\bibfnamefont {I.}~\bibnamefont
  {Gross}}, \bibinfo {author} {\bibfnamefont {L.}~\bibnamefont
  {Mart{\'\i}nez}}, \bibinfo {author} {\bibfnamefont {J.-P.}\ \bibnamefont
  {Tetienne}}, \bibinfo {author} {\bibfnamefont {T.}~\bibnamefont {Hingant}},
  \bibinfo {author} {\bibfnamefont {J.-F.}\ \bibnamefont {Roch}}, \bibinfo
  {author} {\bibfnamefont {K.}~\bibnamefont {Garcia}}, \bibinfo {author}
  {\bibfnamefont {R.}~\bibnamefont {Soucaille}}, \bibinfo {author}
  {\bibfnamefont {J.}~\bibnamefont {Adam}}, \bibinfo {author} {\bibfnamefont
  {J.-V.}\ \bibnamefont {Kim}}, \bibinfo {author} {\bibfnamefont
  {S.}~\bibnamefont {Rohart}}, \emph {et~al.},\ }\bibfield  {title} {\bibinfo
  {title} {Direct measurement of interfacial dzyaloshinskii-moriya interaction
  in x| cofeb| mgo heterostructures with a scanning nv magnetometer (x= ta,
  tan, and w)},\ }\href@noop {} {\bibfield  {journal} {\bibinfo  {journal}
  {Physical Review B}\ }\textbf {\bibinfo {volume} {94}},\ \bibinfo {pages}
  {064413} (\bibinfo {year} {2016})}\BibitemShut {NoStop}%
\bibitem [{\citenamefont {Turov}(1996)}]{turov1996symmetry}%
  \BibitemOpen
  \bibfield  {author} {\bibinfo {author} {\bibfnamefont {E.}~\bibnamefont
  {Turov}},\ }\bibfield  {title} {\bibinfo {title} {Symmetry and physical
  properties of antiferromagnetic macrostructures},\ }\href@noop {} {\bibfield
  {journal} {\bibinfo  {journal} {EPL (Europhysics Letters)}\ }\textbf
  {\bibinfo {volume} {34}},\ \bibinfo {pages} {305} (\bibinfo {year}
  {1996})}\BibitemShut {NoStop}%
\bibitem [{\citenamefont {Drchal}\ \emph {et~al.}(1996)\citenamefont {Drchal},
  \citenamefont {Kudrnovsk{\`y}}, \citenamefont {Turek},\ and\ \citenamefont
  {Weinberger}}]{drchal1996interlayer}%
  \BibitemOpen
  \bibfield  {author} {\bibinfo {author} {\bibfnamefont {V.}~\bibnamefont
  {Drchal}}, \bibinfo {author} {\bibfnamefont {J.}~\bibnamefont
  {Kudrnovsk{\`y}}}, \bibinfo {author} {\bibfnamefont {I.}~\bibnamefont
  {Turek}},\ and\ \bibinfo {author} {\bibfnamefont {P.}~\bibnamefont
  {Weinberger}},\ }\bibfield  {title} {\bibinfo {title} {Interlayer magnetic
  coupling: The torque method},\ }\href@noop {} {\bibfield  {journal} {\bibinfo
   {journal} {Physical Review B}\ }\textbf {\bibinfo {volume} {53}},\ \bibinfo
  {pages} {15036} (\bibinfo {year} {1996})}\BibitemShut {NoStop}%
\bibitem [{\citenamefont {Alamdar}\ \emph {et~al.}(2021)\citenamefont
  {Alamdar}, \citenamefont {Leonard}, \citenamefont {Cui}, \citenamefont
  {Rimal}, \citenamefont {Xue}, \citenamefont {Akinola}, \citenamefont
  {Patrick~Xiao}, \citenamefont {Friedman}, \citenamefont {Bennett},
  \citenamefont {Marinella} \emph {et~al.}}]{alamdar2021domain}%
  \BibitemOpen
  \bibfield  {author} {\bibinfo {author} {\bibfnamefont {M.}~\bibnamefont
  {Alamdar}}, \bibinfo {author} {\bibfnamefont {T.}~\bibnamefont {Leonard}},
  \bibinfo {author} {\bibfnamefont {C.}~\bibnamefont {Cui}}, \bibinfo {author}
  {\bibfnamefont {B.~P.}\ \bibnamefont {Rimal}}, \bibinfo {author}
  {\bibfnamefont {L.}~\bibnamefont {Xue}}, \bibinfo {author} {\bibfnamefont
  {O.~G.}\ \bibnamefont {Akinola}}, \bibinfo {author} {\bibfnamefont
  {T.}~\bibnamefont {Patrick~Xiao}}, \bibinfo {author} {\bibfnamefont {J.~S.}\
  \bibnamefont {Friedman}}, \bibinfo {author} {\bibfnamefont {C.~H.}\
  \bibnamefont {Bennett}}, \bibinfo {author} {\bibfnamefont {M.~J.}\
  \bibnamefont {Marinella}}, \emph {et~al.},\ }\bibfield  {title} {\bibinfo
  {title} {Domain wall-magnetic tunnel junction spin--orbit torque devices and
  circuits for in-memory computing},\ }\href@noop {} {\bibfield  {journal}
  {\bibinfo  {journal} {Applied Physics Letters}\ }\textbf {\bibinfo {volume}
  {118}},\ \bibinfo {pages} {112401} (\bibinfo {year} {2021})}\BibitemShut
  {NoStop}%
\bibitem [{\citenamefont {Bogdanov}\ and\ \citenamefont
  {R{\"o}{\ss}ler}(2001)}]{bogdanov2001chiral}%
  \BibitemOpen
  \bibfield  {author} {\bibinfo {author} {\bibfnamefont {A.}~\bibnamefont
  {Bogdanov}}\ and\ \bibinfo {author} {\bibfnamefont {U.}~\bibnamefont
  {R{\"o}{\ss}ler}},\ }\bibfield  {title} {\bibinfo {title} {Chiral symmetry
  breaking in magnetic thin films and multilayers},\ }\href@noop {} {\bibfield
  {journal} {\bibinfo  {journal} {Physical review letters}\ }\textbf {\bibinfo
  {volume} {87}},\ \bibinfo {pages} {037203} (\bibinfo {year}
  {2001})}\BibitemShut {NoStop}%
\bibitem [{\citenamefont {Yang}\ \emph {et~al.}(2015)\citenamefont {Yang},
  \citenamefont {Thiaville}, \citenamefont {Rohart}, \citenamefont {Fert},\
  and\ \citenamefont {Chshiev}}]{yang2015anatomy}%
  \BibitemOpen
  \bibfield  {author} {\bibinfo {author} {\bibfnamefont {H.}~\bibnamefont
  {Yang}}, \bibinfo {author} {\bibfnamefont {A.}~\bibnamefont {Thiaville}},
  \bibinfo {author} {\bibfnamefont {S.}~\bibnamefont {Rohart}}, \bibinfo
  {author} {\bibfnamefont {A.}~\bibnamefont {Fert}},\ and\ \bibinfo {author}
  {\bibfnamefont {M.}~\bibnamefont {Chshiev}},\ }\bibfield  {title} {\bibinfo
  {title} {Anatomy of dzyaloshinskii-moriya interaction at co/pt interfaces},\
  }\href@noop {} {\bibfield  {journal} {\bibinfo  {journal} {Physical review
  letters}\ }\textbf {\bibinfo {volume} {115}},\ \bibinfo {pages} {267210}
  (\bibinfo {year} {2015})}\BibitemShut {NoStop}%
\bibitem [{\citenamefont {Belabbes}\ \emph {et~al.}(2016)\citenamefont
  {Belabbes}, \citenamefont {Bihlmayer}, \citenamefont {Bechstedt},
  \citenamefont {Bl{\"u}gel},\ and\ \citenamefont
  {Manchon}}]{belabbes2016hund}%
  \BibitemOpen
  \bibfield  {author} {\bibinfo {author} {\bibfnamefont {A.}~\bibnamefont
  {Belabbes}}, \bibinfo {author} {\bibfnamefont {G.}~\bibnamefont {Bihlmayer}},
  \bibinfo {author} {\bibfnamefont {F.}~\bibnamefont {Bechstedt}}, \bibinfo
  {author} {\bibfnamefont {S.}~\bibnamefont {Bl{\"u}gel}},\ and\ \bibinfo
  {author} {\bibfnamefont {A.}~\bibnamefont {Manchon}},\ }\bibfield  {title}
  {\bibinfo {title} {Hund’s rule-driven dzyaloshinskii-moriya interaction at
  3 d- 5 d interfaces},\ }\href@noop {} {\bibfield  {journal} {\bibinfo
  {journal} {Physical review letters}\ }\textbf {\bibinfo {volume} {117}},\
  \bibinfo {pages} {247202} (\bibinfo {year} {2016})}\BibitemShut {NoStop}%
\bibitem [{\citenamefont {Avci}\ \emph {et~al.}(2021)\citenamefont {Avci},
  \citenamefont {Lambert}, \citenamefont {Sala},\ and\ \citenamefont
  {Gambardella}}]{avci2021chiral}%
  \BibitemOpen
  \bibfield  {author} {\bibinfo {author} {\bibfnamefont {C.~O.}\ \bibnamefont
  {Avci}}, \bibinfo {author} {\bibfnamefont {C.-H.}\ \bibnamefont {Lambert}},
  \bibinfo {author} {\bibfnamefont {G.}~\bibnamefont {Sala}},\ and\ \bibinfo
  {author} {\bibfnamefont {P.}~\bibnamefont {Gambardella}},\ }\bibfield
  {title} {\bibinfo {title} {Chiral coupling between magnetic layers with
  orthogonal magnetization},\ }\href@noop {} {\bibfield  {journal} {\bibinfo
  {journal} {Physical review letters}\ }\textbf {\bibinfo {volume} {127}},\
  \bibinfo {pages} {167202} (\bibinfo {year} {2021})}\BibitemShut {NoStop}%
\bibitem [{\citenamefont {Liu}\ \emph {et~al.}(2021)\citenamefont {Liu},
  \citenamefont {Luo}, \citenamefont {Rohart}, \citenamefont {Heyderman},
  \citenamefont {Gambardella},\ and\ \citenamefont
  {Hrabec}}]{liu2021engineering}%
  \BibitemOpen
  \bibfield  {author} {\bibinfo {author} {\bibfnamefont {Z.}~\bibnamefont
  {Liu}}, \bibinfo {author} {\bibfnamefont {Z.}~\bibnamefont {Luo}}, \bibinfo
  {author} {\bibfnamefont {S.}~\bibnamefont {Rohart}}, \bibinfo {author}
  {\bibfnamefont {L.~J.}\ \bibnamefont {Heyderman}}, \bibinfo {author}
  {\bibfnamefont {P.}~\bibnamefont {Gambardella}},\ and\ \bibinfo {author}
  {\bibfnamefont {A.}~\bibnamefont {Hrabec}},\ }\bibfield  {title} {\bibinfo
  {title} {Engineering of intrinsic chiral torques in magnetic thin films based
  on the dzyaloshinskii-moriya interaction},\ }\href@noop {} {\bibfield
  {journal} {\bibinfo  {journal} {Physical Review Applied}\ }\textbf {\bibinfo
  {volume} {16}},\ \bibinfo {pages} {054049} (\bibinfo {year}
  {2021})}\BibitemShut {NoStop}%
\bibitem [{\citenamefont {Gr\"unberg}\ \emph {et~al.}(1986)\citenamefont
  {Gr\"unberg}, \citenamefont {Schreiber}, \citenamefont {Pang}, \citenamefont
  {Brodsky},\ and\ \citenamefont {Sowers}}]{PhysRevLett.57.2442}%
  \BibitemOpen
  \bibfield  {author} {\bibinfo {author} {\bibfnamefont {P.}~\bibnamefont
  {Gr\"unberg}}, \bibinfo {author} {\bibfnamefont {R.}~\bibnamefont
  {Schreiber}}, \bibinfo {author} {\bibfnamefont {Y.}~\bibnamefont {Pang}},
  \bibinfo {author} {\bibfnamefont {M.~B.}\ \bibnamefont {Brodsky}},\ and\
  \bibinfo {author} {\bibfnamefont {H.}~\bibnamefont {Sowers}},\ }\bibfield
  {title} {\bibinfo {title} {Layered magnetic structures: Evidence for
  antiferromagnetic coupling of fe layers across cr interlayers},\ }\href@noop
  {} {\bibfield  {journal} {\bibinfo  {journal} {Phys. Rev. Lett.}\ }\textbf
  {\bibinfo {volume} {57}},\ \bibinfo {pages} {2442} (\bibinfo {year}
  {1986})}\BibitemShut {NoStop}%
\bibitem [{\citenamefont {Lau}\ \emph {et~al.}(2016)\citenamefont {Lau},
  \citenamefont {Betto}, \citenamefont {Rode}, \citenamefont {Coey},\ and\
  \citenamefont {Stamenov}}]{lau2016spin}%
  \BibitemOpen
  \bibfield  {author} {\bibinfo {author} {\bibfnamefont {Y.-C.}\ \bibnamefont
  {Lau}}, \bibinfo {author} {\bibfnamefont {D.}~\bibnamefont {Betto}}, \bibinfo
  {author} {\bibfnamefont {K.}~\bibnamefont {Rode}}, \bibinfo {author}
  {\bibfnamefont {J.}~\bibnamefont {Coey}},\ and\ \bibinfo {author}
  {\bibfnamefont {P.}~\bibnamefont {Stamenov}},\ }\bibfield  {title} {\bibinfo
  {title} {Spin--orbit torque switching without an external field using
  interlayer exchange coupling},\ }\href@noop {} {\bibfield  {journal}
  {\bibinfo  {journal} {Nature nanotechnology}\ }\textbf {\bibinfo {volume}
  {11}},\ \bibinfo {pages} {758} (\bibinfo {year} {2016})}\BibitemShut
  {NoStop}%
\bibitem [{\citenamefont {Avci}\ \emph {et~al.}(2019)\citenamefont {Avci},
  \citenamefont {Rosenberg}, \citenamefont {Caretta}, \citenamefont
  {B{\"u}ttner}, \citenamefont {Mann}, \citenamefont {Marcus}, \citenamefont
  {Bono}, \citenamefont {Ross},\ and\ \citenamefont
  {Beach}}]{avci2019interface}%
  \BibitemOpen
  \bibfield  {author} {\bibinfo {author} {\bibfnamefont {C.~O.}\ \bibnamefont
  {Avci}}, \bibinfo {author} {\bibfnamefont {E.}~\bibnamefont {Rosenberg}},
  \bibinfo {author} {\bibfnamefont {L.}~\bibnamefont {Caretta}}, \bibinfo
  {author} {\bibfnamefont {F.}~\bibnamefont {B{\"u}ttner}}, \bibinfo {author}
  {\bibfnamefont {M.}~\bibnamefont {Mann}}, \bibinfo {author} {\bibfnamefont
  {C.}~\bibnamefont {Marcus}}, \bibinfo {author} {\bibfnamefont
  {D.}~\bibnamefont {Bono}}, \bibinfo {author} {\bibfnamefont {C.~A.}\
  \bibnamefont {Ross}},\ and\ \bibinfo {author} {\bibfnamefont {G.~S.}\
  \bibnamefont {Beach}},\ }\bibfield  {title} {\bibinfo {title}
  {Interface-driven chiral magnetism and current-driven domain walls in
  insulating magnetic garnets},\ }\href@noop {} {\bibfield  {journal} {\bibinfo
   {journal} {Nature nanotechnology}\ }\textbf {\bibinfo {volume} {14}},\
  \bibinfo {pages} {561} (\bibinfo {year} {2019})}\BibitemShut {NoStop}%
\bibitem [{\citenamefont {Manchon}\ and\ \citenamefont
  {Zhang}(2009)}]{manchon2009theory}%
  \BibitemOpen
  \bibfield  {author} {\bibinfo {author} {\bibfnamefont {A.}~\bibnamefont
  {Manchon}}\ and\ \bibinfo {author} {\bibfnamefont {S.}~\bibnamefont
  {Zhang}},\ }\bibfield  {title} {\bibinfo {title} {Theory of spin torque due
  to spin-orbit coupling},\ }\href@noop {} {\bibfield  {journal} {\bibinfo
  {journal} {Physical Review B}\ }\textbf {\bibinfo {volume} {79}},\ \bibinfo
  {pages} {094422} (\bibinfo {year} {2009})}\BibitemShut {NoStop}%
\bibitem [{\citenamefont {Caretta}\ \emph {et~al.}(2020)\citenamefont
  {Caretta}, \citenamefont {Rosenberg}, \citenamefont {B{\"u}ttner},
  \citenamefont {Fakhrul}, \citenamefont {Gargiani}, \citenamefont
  {Valvidares}, \citenamefont {Chen}, \citenamefont {Reddy}, \citenamefont
  {Muller}, \citenamefont {Ross} \emph {et~al.}}]{caretta2020interfacial}%
  \BibitemOpen
  \bibfield  {author} {\bibinfo {author} {\bibfnamefont {L.}~\bibnamefont
  {Caretta}}, \bibinfo {author} {\bibfnamefont {E.}~\bibnamefont {Rosenberg}},
  \bibinfo {author} {\bibfnamefont {F.}~\bibnamefont {B{\"u}ttner}}, \bibinfo
  {author} {\bibfnamefont {T.}~\bibnamefont {Fakhrul}}, \bibinfo {author}
  {\bibfnamefont {P.}~\bibnamefont {Gargiani}}, \bibinfo {author}
  {\bibfnamefont {M.}~\bibnamefont {Valvidares}}, \bibinfo {author}
  {\bibfnamefont {Z.}~\bibnamefont {Chen}}, \bibinfo {author} {\bibfnamefont
  {P.}~\bibnamefont {Reddy}}, \bibinfo {author} {\bibfnamefont {D.~A.}\
  \bibnamefont {Muller}}, \bibinfo {author} {\bibfnamefont {C.~A.}\
  \bibnamefont {Ross}}, \emph {et~al.},\ }\bibfield  {title} {\bibinfo {title}
  {Interfacial dzyaloshinskii-moriya interaction arising from rare-earth
  orbital magnetism in insulating magnetic oxides},\ }\href@noop {} {\bibfield
  {journal} {\bibinfo  {journal} {Nature communications}\ }\textbf {\bibinfo
  {volume} {11}},\ \bibinfo {pages} {1} (\bibinfo {year} {2020})}\BibitemShut
  {NoStop}%
\bibitem [{\citenamefont {Brock}\ \emph {et~al.}(2021)\citenamefont {Brock},
  \citenamefont {Kitcher}, \citenamefont {Vallobra}, \citenamefont {Medapalli},
  \citenamefont {Li}, \citenamefont {De~Graef}, \citenamefont {Riley},
  \citenamefont {Nembach}, \citenamefont {Mangin}, \citenamefont {Sokalski}
  \emph {et~al.}}]{brock2021dynamic}%
  \BibitemOpen
  \bibfield  {author} {\bibinfo {author} {\bibfnamefont {J.~A.}\ \bibnamefont
  {Brock}}, \bibinfo {author} {\bibfnamefont {M.~D.}\ \bibnamefont {Kitcher}},
  \bibinfo {author} {\bibfnamefont {P.}~\bibnamefont {Vallobra}}, \bibinfo
  {author} {\bibfnamefont {R.}~\bibnamefont {Medapalli}}, \bibinfo {author}
  {\bibfnamefont {M.~P.}\ \bibnamefont {Li}}, \bibinfo {author} {\bibfnamefont
  {M.}~\bibnamefont {De~Graef}}, \bibinfo {author} {\bibfnamefont {G.~A.}\
  \bibnamefont {Riley}}, \bibinfo {author} {\bibfnamefont {H.~T.}\ \bibnamefont
  {Nembach}}, \bibinfo {author} {\bibfnamefont {S.}~\bibnamefont {Mangin}},
  \bibinfo {author} {\bibfnamefont {V.}~\bibnamefont {Sokalski}}, \emph
  {et~al.},\ }\bibfield  {title} {\bibinfo {title} {Dynamic symmetry breaking
  in chiral magnetic systems},\ }\href@noop {} {\bibfield  {journal} {\bibinfo
  {journal} {Advanced Materials}\ }\textbf {\bibinfo {volume} {33}},\ \bibinfo
  {pages} {2101524} (\bibinfo {year} {2021})}\BibitemShut {NoStop}%
\bibitem [{\citenamefont {Ding}\ \emph {et~al.}(2019)\citenamefont {Ding},
  \citenamefont {Ross}, \citenamefont {Lebrun}, \citenamefont {Becker},
  \citenamefont {Lee}, \citenamefont {Boventer}, \citenamefont {Das},
  \citenamefont {Kurokawa}, \citenamefont {Gupta}, \citenamefont {Yang} \emph
  {et~al.}}]{ding2019interfacial}%
  \BibitemOpen
  \bibfield  {author} {\bibinfo {author} {\bibfnamefont {S.}~\bibnamefont
  {Ding}}, \bibinfo {author} {\bibfnamefont {A.}~\bibnamefont {Ross}}, \bibinfo
  {author} {\bibfnamefont {R.}~\bibnamefont {Lebrun}}, \bibinfo {author}
  {\bibfnamefont {S.}~\bibnamefont {Becker}}, \bibinfo {author} {\bibfnamefont
  {K.}~\bibnamefont {Lee}}, \bibinfo {author} {\bibfnamefont {I.}~\bibnamefont
  {Boventer}}, \bibinfo {author} {\bibfnamefont {S.}~\bibnamefont {Das}},
  \bibinfo {author} {\bibfnamefont {Y.}~\bibnamefont {Kurokawa}}, \bibinfo
  {author} {\bibfnamefont {S.}~\bibnamefont {Gupta}}, \bibinfo {author}
  {\bibfnamefont {J.}~\bibnamefont {Yang}}, \emph {et~al.},\ }\bibfield
  {title} {\bibinfo {title} {Interfacial dzyaloshinskii-moriya interaction and
  chiral magnetic textures in a ferrimagnetic insulator},\ }\href@noop {}
  {\bibfield  {journal} {\bibinfo  {journal} {Physical Review B}\ }\textbf
  {\bibinfo {volume} {100}},\ \bibinfo {pages} {100406} (\bibinfo {year}
  {2019})}\BibitemShut {NoStop}%
\bibitem [{sup()}]{supp}%
  \BibitemOpen
  \href@noop {} {}\bibinfo {note} {See supplemental material for
  details}\BibitemShut {NoStop}%
\bibitem [{\citenamefont {Antropov}\ \emph {et~al.}(2021)\citenamefont
  {Antropov}, \citenamefont {Kravtsov}, \citenamefont {Makarova}, \citenamefont
  {Proglyado}, \citenamefont {Keller}, \citenamefont {Subbotin}, \citenamefont
  {Pashaev}, \citenamefont {Prutskov}, \citenamefont {Vasiliev}, \citenamefont
  {Chesnokov} \emph {et~al.}}]{antropov2021tunable}%
  \BibitemOpen
  \bibfield  {author} {\bibinfo {author} {\bibfnamefont {N.}~\bibnamefont
  {Antropov}}, \bibinfo {author} {\bibfnamefont {E.}~\bibnamefont {Kravtsov}},
  \bibinfo {author} {\bibfnamefont {M.}~\bibnamefont {Makarova}}, \bibinfo
  {author} {\bibfnamefont {V.}~\bibnamefont {Proglyado}}, \bibinfo {author}
  {\bibfnamefont {T.}~\bibnamefont {Keller}}, \bibinfo {author} {\bibfnamefont
  {I.}~\bibnamefont {Subbotin}}, \bibinfo {author} {\bibfnamefont
  {E.}~\bibnamefont {Pashaev}}, \bibinfo {author} {\bibfnamefont
  {G.}~\bibnamefont {Prutskov}}, \bibinfo {author} {\bibfnamefont
  {A.}~\bibnamefont {Vasiliev}}, \bibinfo {author} {\bibfnamefont {Y.~M.}\
  \bibnamefont {Chesnokov}}, \emph {et~al.},\ }\bibfield  {title} {\bibinfo
  {title} {Tunable spin-flop transition in artificial ferrimagnets},\
  }\href@noop {} {\bibfield  {journal} {\bibinfo  {journal} {Physical Review
  B}\ }\textbf {\bibinfo {volume} {104}},\ \bibinfo {pages} {054414} (\bibinfo
  {year} {2021})}\BibitemShut {NoStop}%
\bibitem [{\citenamefont {Welp}\ \emph {et~al.}(1999)\citenamefont {Welp},
  \citenamefont {Berger}, \citenamefont {Miller}, \citenamefont
  {Vlasko-Vlasov}, \citenamefont {Gray},\ and\ \citenamefont
  {Mitchell}}]{welp1999direct}%
  \BibitemOpen
  \bibfield  {author} {\bibinfo {author} {\bibfnamefont {U.}~\bibnamefont
  {Welp}}, \bibinfo {author} {\bibfnamefont {A.}~\bibnamefont {Berger}},
  \bibinfo {author} {\bibfnamefont {D.}~\bibnamefont {Miller}}, \bibinfo
  {author} {\bibfnamefont {V.}~\bibnamefont {Vlasko-Vlasov}}, \bibinfo {author}
  {\bibfnamefont {K.}~\bibnamefont {Gray}},\ and\ \bibinfo {author}
  {\bibfnamefont {J.}~\bibnamefont {Mitchell}},\ }\bibfield  {title} {\bibinfo
  {title} {Direct imaging of the first-order spin-flop transition in the
  layered manganite la 1.4 sr 1.6 mn 2 o 7},\ }\href@noop {} {\bibfield
  {journal} {\bibinfo  {journal} {Physical review letters}\ }\textbf {\bibinfo
  {volume} {83}},\ \bibinfo {pages} {4180} (\bibinfo {year}
  {1999})}\BibitemShut {NoStop}%
\bibitem [{\citenamefont {Katsura}\ \emph {et~al.}(2005)\citenamefont
  {Katsura}, \citenamefont {Nagaosa},\ and\ \citenamefont
  {Balatsky}}]{katsura2005}%
  \BibitemOpen
  \bibfield  {author} {\bibinfo {author} {\bibfnamefont {H.}~\bibnamefont
  {Katsura}}, \bibinfo {author} {\bibfnamefont {N.}~\bibnamefont {Nagaosa}},\
  and\ \bibinfo {author} {\bibfnamefont {A.~V.}\ \bibnamefont {Balatsky}},\
  }\bibfield  {title} {\bibinfo {title} {Spin current and magnetoelectric
  effect in noncollinear magnets},\ }\href@noop {} {\bibfield  {journal}
  {\bibinfo  {journal} {Physical review letters}\ }\textbf {\bibinfo {volume}
  {95}},\ \bibinfo {pages} {057205} (\bibinfo {year} {2005})}\BibitemShut
  {NoStop}%
\bibitem [{\citenamefont {Galkina}\ and\ \citenamefont
  {Ivanov}(2018)}]{galkina2018dynamic}%
  \BibitemOpen
  \bibfield  {author} {\bibinfo {author} {\bibfnamefont {E.}~\bibnamefont
  {Galkina}}\ and\ \bibinfo {author} {\bibfnamefont {B.}~\bibnamefont
  {Ivanov}},\ }\bibfield  {title} {\bibinfo {title} {Dynamic solitons in
  antiferromagnets},\ }\href@noop {} {\bibfield  {journal} {\bibinfo  {journal}
  {Low Temperature Physics}\ }\textbf {\bibinfo {volume} {44}},\ \bibinfo
  {pages} {618} (\bibinfo {year} {2018})}\BibitemShut {NoStop}%
\bibitem [{\citenamefont {Hone}\ and\ \citenamefont
  {Leung}(1980)}]{hone1980spin}%
  \BibitemOpen
  \bibfield  {author} {\bibinfo {author} {\bibfnamefont {D.}~\bibnamefont
  {Hone}}\ and\ \bibinfo {author} {\bibfnamefont {K.}~\bibnamefont {Leung}},\
  }\bibfield  {title} {\bibinfo {title} {Spin-correlation functions in
  sine-gordon magnetic chains},\ }\href@noop {} {\bibfield  {journal} {\bibinfo
   {journal} {Physical Review B}\ }\textbf {\bibinfo {volume} {22}},\ \bibinfo
  {pages} {5308} (\bibinfo {year} {1980})}\BibitemShut {NoStop}%
\bibitem [{\citenamefont {Cuevas-Maraver}\ \emph {et~al.}(2014)\citenamefont
  {Cuevas-Maraver}, \citenamefont {Kevrekidis},\ and\ \citenamefont
  {Williams}}]{cuevas2014sine}%
  \BibitemOpen
  \bibfield  {author} {\bibinfo {author} {\bibfnamefont {J.}~\bibnamefont
  {Cuevas-Maraver}}, \bibinfo {author} {\bibfnamefont {P.~G.}\ \bibnamefont
  {Kevrekidis}},\ and\ \bibinfo {author} {\bibfnamefont {F.}~\bibnamefont
  {Williams}},\ }\bibfield  {title} {\bibinfo {title} {The sine-gordon model
  and its applications},\ }\href@noop {} {\bibfield  {journal} {\bibinfo
  {journal} {Nonlinear systems and complexity}\ }\textbf {\bibinfo {volume}
  {10}} (\bibinfo {year} {2014})}\BibitemShut {NoStop}%
\bibitem [{\citenamefont {Mellado}\ \emph {et~al.}(2012)\citenamefont
  {Mellado}, \citenamefont {Concha},\ and\ \citenamefont
  {Mahadevan}}]{mellado2012macroscopic}%
  \BibitemOpen
  \bibfield  {author} {\bibinfo {author} {\bibfnamefont {P.}~\bibnamefont
  {Mellado}}, \bibinfo {author} {\bibfnamefont {A.}~\bibnamefont {Concha}},\
  and\ \bibinfo {author} {\bibfnamefont {L.}~\bibnamefont {Mahadevan}},\
  }\bibfield  {title} {\bibinfo {title} {Macroscopic magnetic frustration},\
  }\href@noop {} {\bibfield  {journal} {\bibinfo  {journal} {Physical review
  letters}\ }\textbf {\bibinfo {volume} {109}},\ \bibinfo {pages} {257203}
  (\bibinfo {year} {2012})}\BibitemShut {NoStop}%
\end{thebibliography}
%

\end{document}